\documentclass[english]{article}
\usepackage[OT1]{fontenc}
\usepackage[latin9]{inputenc}
\usepackage[dvips]{graphicx}
\usepackage{epsfig}
\usepackage{subfigure}
\usepackage{amsmath, amssymb}

\newlength{\wth}
\def\Ptmiss{\not{\hbox{\kern-3pt $E_T$}}}

\makeatletter

\newcommand{\mathsym}[1]{{}}

\usepackage{geometry}



\usepackage{babel}
\makeatother

\begin{document}

\title{Identifying Multi-Top Events from Gluino Decay at the LHC}

\author{B. S. Acharya$^1$, P. Grajek$^2$, G. L. Kane$^2$, E. Kuflik$^2$,
  K. Suruliz$^1$,
  Lian-Tao Wang$^3$ \\
\textit{$^1$ Abdus Salam International Center for Theoretical Physics
} \\
\textit{Strada Costiera 11,34014 Trieste. ITALY and }
\textit{INFN, Sezione di Trieste.} \\
\textit{$^2$  Michigan Center for Theoretical Physics,} \\ \textit{University of
  Michigan,}   \textit{Ann Arbor, MI 48109, USA.} \\
\textit{$^3$ Department of Physics,} \textit{ Princeton University,}
  \textit{Princeton, NJ 08540, USA}}

\maketitle
\begin{abstract}
We study the LHC signal of a light gluino whose cascade
decay is dominated by channels involving top, and, sometimes, bottom
quarks. This is a generic signature for a number of
supersymmetry breaking scenarios considered recently, where the
squarks are heavier than gauginos. Third generation final
states generically dominate since third generation squarks are
typically somewhat lighter in these models. At the LHC we demonstrate 
that early discovery is possible due to the existence
of multi-lepton multi-bottom final states which have fairly low
Standard Model background. We find that the best discovery channel is
'same sign dilepton'.
The relative decay branching ratios into $tt$, $tb$ and $bb$ states carry
important information about the underlying model.  Although
reconstruction will yield evidence for
the existence of top quarks in the event, we demonstrate that
identifying multiple top quarks suffers from low efficiency and
large combinatorial background, due to
the large number of final state particles. We propose a fitting method
which takes advantage of excesses in a large number of
channels. We demonstrate such a method will allow us to extract
information about decay branching ratios with moderate integrated
luminosities. In addition, the method also gives
an upper bound on the gluino production cross
section and an estimate of the gluino mass.

\end{abstract}

\newpage

\section{Introduction}

\label{sec:introduction}

Many considerations of new physics at the TeV scale point toward top
rich final states at the LHC. Such scenarios include top compositeness
\cite{top-composite}, and models in which top partners ensure the 
naturalness of electroweak
symmetry breaking \cite{susy,ued,lh,twin}. Compared with the Standard Model
QCD production
of $t\bar{t}$, top quarks in the final states of new physics production
typically either have very different kinematics
\cite{resonance-ttbar,Brooijmans,Bai:2008sk}, or different event topology
\cite{tprime-ttet,Kraml:2005kb,Lillie:2007hd,Contino:2008hi,Han:2008xb},
which may make it crucial to develop new techniques to identify them.

Naturalness of the electroweak symmetry breaking typically requires
the existence of a light top partner. Experimental
observations, ranging from the existence of Cold Dark Matter in the
universe to the constraints of electroweak precision measurements,
strongly motivates the existence of a neutral stable particle as particle
of new physics signals at the LHC. The above considerations lead to
the recent studies of new physics signals with $t\bar{t}+{\not}E_{T}$
final state \cite{tprime-ttet}.

In this article, we consider scenarios which also include a gluon
partner,  such as gluino in
low energy supersymmetry \cite{susy,Chung:2003fi}, the KK-gluon in
universal extra dimension models \cite{ued}, or other octet states
\cite{Gerbush:2007fe,scalar-octet,Dobrescu:2007yp}. Due to the nature of
proton colliders, production of such color
octet gluon partners and their decay typically becomes the main channel
of new physics signals. Decay products of the gluon partner typically
include an even number of quarks. Combining this with
the scenario with light top partners, we conclude that a typical signature
of production of gluon partners will be multiple top quarks in the
final states.

In the rest of this paper, we will study this signature using a particular
example, low energy supersymmetry with light gluinos. We will focus on the
scenario in which the squarks are heavier than the gluino and the
third generation squarks are lighter than those of the first two
generations. In this case,  the gluino will dominantly decay into top
(bottom) quarks. Although not absolutely
unavoidable, this is a generic possibility from the point of view
of many models being studied recently
\cite{mirage,zprime-med,Acharya:2008zi,Heckman:2008rb}. Heavier squark
masses are often preferred due to
  constraints from flavor changing neutral currents
  \cite{eff_susy,focus_point,split}.  RGE running of scalar
masses from the high scale down to the electroweak scale will tend to push
the third generation squark masses significantly lower than those of
the other generations. Large third generation trilinear couplings will
also help further lower one of the stop masses. For earlier studies on
gluino decay into third generation quarks, see Refs.
~\cite{Baer:1990sc,Hisano:2002xq,Hisano:2003qu,Mercadante:2007zz,Baer:2007ya,Gambino:2005eh,Toharia:2005gm}.

In the minimal case, this scenario has only a single, light top partner,
$\tilde{t}_{R}$. On the other hand, we would like to include in our study the
possibility that all the squarks of the full third generation could be light
relative to the other two generations. Therefore, we are led to consider decay
channels $\tilde{g}\rightarrow t\bar{t}\tilde{N}$, $\tilde{g}\rightarrow
t\bar{b}\tilde{C}^{-}$, and $\tilde{g}\rightarrow b\bar{b}\tilde{N}$. As will be
clear from our discussion later, many of the leading order features of the
signature of different combinations of the above decay channels (from gluino
pair production and decay) can be quite similar. Therefore, it is one of the
primary purposes of this article to study techniques to distinguish them. We
remark that such measurement is crucial for understanding both the spectrum of
the third generation sfermions and the electroweak-inos. For example, if we
measure a non-zero branching ratio for the decay channel $\tilde{g}\rightarrow
t\bar{b}\tilde{\chi}^{-}$, then there must be a light electroweak-ino carrying
charge, suggesting that the lightest supersymmetric particle (LSP) is 
mostly
Wino. Moreover, the relative branching ratio $BR(\tilde{g} \rightarrow t \bar{t}
+ \chi)/BR(\tilde{g} \rightarrow b \bar{b} + \chi) $ carries important
information about the squark masses. For example, if this ratio is close to 1,
it strongly suggests that left-handed squark masses are lighter than the
right-handed ones, $m_{\tilde{Q}_3} < m_{\tilde{t}_R}, \ m_{\tilde{b}_R}$.

In section \ref{sec:Signal-Isolation}, we will focus on discovering
new physics in this class of final states.
Decay of multiple top quarks could lead to b-rich and lepton rich final
states. Therefore, we expect great potential for early discovery. For
example, we show that significant excesses can be observed in many
channels even with just 500 pb$^{-1}$ of data. The obvious channel
with the best early discovery potential is same-sign dilepton plus
additional b-tags.

In section \ref{sec:reconstruction}, we study the problem of direct
reconstruction of top quarks. Large combinatorics and high probability of object
merging are expected due to the large multiplicity of final state particles.
Therefore, while we may gather evidence from this study that there are
\textit{some} top quarks in the decay chain of the gluino, similar to the
approaches taken in Ref.~\cite{Hisano:2002xq,Hisano:2003qu,Gerbush:2007fe}, the
direct reconstruction efficiency for the top quark is low. We find that while
the efficiency for reconstructing a single top quark candidate approaches
$49\%$, the efficiency to detect three and four candidates drops dramatically to
$1.5\%$ and $0.02\%$, respectively. As a result, it is less likely we can
measure top multiplicity by direct reconstruction.

We will demonstrate in our study a fitting procedure which could allow
us to measure the branching ratios of different gluino decay channels.
We begin by simulating a number of samples of gluino pair production and
decay, each with different final states, such as $tttt$, $tttb$,
$ttbb$, and so on. Then we will fit the relative weights of different
samples to match a set of experimental signatures. Of course, without
precise knowledge of the underlying spectrum, choice of the templates
will introduce errors in the estimate of the branching ratios. We
studied such effects by using several templates with different hypotheses
for the relevant masses. We conclude that such a method will allow us to
establish important features of gluino branching ratios.

We carry out our study on several benchmark models with relatively low
gluino masses. A detailed scan of the parameter space involving the
gluino mass and different branching ratios is beyond the scope
of this paper. The corresponding results for heavier gluino masses
(but with similar decay branching ratio and mass difference between
gluino and the LSP) could be roughly obtained
by scaling from the present result using relative production cross
sections. The mass gap between the gluino and the neutralino or chargino
in the next step of the decay chain could also have important effects
as it will affect the detection efficiency of various decay products.
In general, a larger mass gap will enhance the discovery potential.
At the same time, we expect this effect is milder in comparison with
the dependence on the gluino mass.

We emphasize that our goal in this study is to demonstrate a method
which allows us to extract information of the SUSY spectrum, such as
the identity of the LSP with relatively low integrated
luminosity. This is possible mainly because, unlike the precision
measurements of the masses and couplings, our method mainly relies on
inclusive counts and general kinematical features. Moreover, since we
do not demand direct reconstruction, we are able to take advantage of
many channels with multiple leptons.  Due to lower background
and theoretical uncertainty in comparison with the pure hadronic
channel, we expect to have significant excesses in many of these
channels. After discovery, we expect our method will yield a first
set of clues about the underlying model during the early stages of LHC
operation.

\section{\label{sec:Physics-Model}Benchmark Models}

\begin{table}
\begin{center}
\begin{tabular}{|c||c|c|c|c|c|c|c||c|c|c|}
\hline
 & \multicolumn{7}{|c||}{Model parameters (TeV) } & 
\multicolumn{3}{c|}{Branching ratios} \\
\hline
 & $m_{\tilde{g}}$ & $m_{\tilde{q}_{1,2}}$ & $m_{\tilde{t}_1}$ &
 $m_{\tilde{t}_2}$ &
$m_{\tilde{b}_1}$ & $m_{\tilde{b}_2}$ & $m_{\tilde{N}, \tilde{C}}$ &
  $(\tilde{g}\rightarrow t\bar{t})$ & $(\tilde{g}\rightarrow
b\bar{b})$ & $(\tilde{g}\rightarrow
tb) $\\
\hline
A & 0.65 & $8$ & $1.3 $ & $8 $ & $2.5$ & $8.1$
& 0.1 &0.92 &0.07 & 0\\
\hline
B & 0.65 & $4$ & 0.8 & 0.93 & 0.87 & $4$ & 0.1 & 0.71 & 0.27 & 0 \\
\hline
C & 0.65 & $4$ & 0.64 & 0.9 & 0.72 & $4$ & 0.1 & 0.52 & 0.47 & 0  \\
\hline
D & 0.65 & $4$ & 0.63 & 0.9 & 0.72  & $4$  & 0.1 & 0.09 & 0.22 & 0.69 \\
\hline
\end{tabular}

\caption{\label{tab:model} Model parameters and relevant 
  branching ratios for the benchmark models considered in this
  paper. The mass parameters are in TeV. The models A, B, and C have
  bino LSP. In Model D, the lightest neutralino and lightest chargino
  are both winos. We have adopted the short
  hand notation where we omit the explicit mention of the identity of
  the electroweak ino in the decay, as it can always be inferred from the
  observable particle content.
  }
\end{center}
\end{table}
We consider four benchmark models in our study. The model parameters and
relevant decay branching ratios are shown in
Table~\ref{tab:model}. For simplicity, we will only consider gluino
decay chains in which the only observable decay products are tops and
bottoms, and they only come from one single step in the gluino decay
chain. This is of course the case if the only state lighter than the
gluino is the neutral LSP, i.e., a bino-like LSP (Models A, B and C). We also
consider the possibility of a wino LSP, in which the neutral LSP is
almost degenerate with the lightest chargino (Model D).

Model A is the simplest example of multi-top physics. It is designed to have
only a single light stop, and therefore always produces four tops in the final
state. Model B and C are designed to include the decay channel $\tilde{g}
\rightarrow b \bar{b} \tilde{N}$, with each model exhibiting a somewhat
different branching ratio. In Model D, the wino LSP is approximately degenerate
with the lightest chargino, which is also wino-like.   It is designed to
further include a chargino in the decay chain. We will study
distinguishing it from the other channels. Since the charged wino
is approximately degenerate with the  wino LSP, it appears only as
missing energy.

In Table~\ref{tab:model} and the rest of this paper, we will adopt a
short-hand notation for gluino decay by only including top and
bottom quarks and not giving explicitly the electroweak-inos, as
it should be evident from the context.

\section{\label{sec:Signal-Isolation}Signal Isolation and Backgrounds}

We begin by discussing the prospects for signal isolation above Standard
Model background at the LHC. The relatively large $b$-jet
and lepton multiplicity associated with four-top production provide
for potentially striking signatures that are easily distinguishable
above the expected SM background. We find that by requesting $\ge3$
$b$-tagged jets and at least one lepton, it is possible to achieve signal
significance $S/\sqrt{B}>3$ for only 500 pb$^{-1}$ of integrated
luminosity. We will demonstrate the discovery potential in three of the four
benchmark models.

One of the important backgrounds from the Standard Model for final states with
many $b$-tagged jets, several isolated leptons and very high missing $E_T$, is
top pair production, $t\bar{t}$. The expected cross-section at the LHC for this
background is $\sigma = 833$ pb (NLO+NLL result
\cite{Bonciani:1998vc}). The $t\bar{t}$ background event samples were
produced using  Pythia 6.4 \cite{Sjostrand:2006za}.

\begin{table}[h!]
\begin{center}
\begin{tabular}{|c|c||c|c||c|c|}
\hline
  Process & $\sigma \left[\mbox{pb}\right]$ & Process & $\sigma
  \left[\mbox{pb}\right]$ & Process & $\sigma \left[\mbox{pb}\right]$
  \\
\hline
 $ t \bar{t}   + 1,2,3 \mbox{ jets} $ & 833 & $ t \bar{b} Z + 1,2
\mbox{ jets}    $ &	0.67 &	$ Z W^+ b	 + 1,2 \mbox{ jets}
$ & 0.48	 \\
\hline
 $ t \bar{t} Z   + 1,2 \mbox{ jets} $ & 0.28 & $ \bar{t} b Z + 1,2
\mbox{ jets}    $ &	0.58 &	$ Z W^- b	 + 1,2 \mbox{ jets}
$ & 0.50	 \\
\hline
 $ t \bar{t} W^- + 1,2 \mbox{ jets} $ & 1.5 & $ t \bar{b} W^+ + 1,2
\mbox{ jets}  $ &	0.18 &	$ Z W^+ \bar{b}	 + 1,2 \mbox{ jets} $
&	0.85  \\
\hline
 $ t \bar{t} W^+ + 1,2 \mbox{ jets} $ & 3.4 & $ \bar{t} b W^- + 1,2
\mbox{ jets}  $ & 0.09 &	$ Z W^- \bar{b}	 + 1,2 \mbox{ jets} $
&	0.28  \\
\hline

\end{tabular}

\caption{\label{tab:SMBCrossSections} Standard Model backgrounds and
  relevant cross sections used.
  }
\end{center}
\end{table}

We have also included in our analysis a set of SM backgrounds involving
associated production of W/Z bosons with third generation quarks. These
contribute significantly to signals with high lepton multiplicity, or same sign
dileptons in the final state. As we will see, the latter case is a particularly
important discovery channel early on. The parton-level SM background event
samples were produced with Madgraph v.4.2.3 \cite{Alwall:2007st}, with the
exception of the $t\bar{t}$ background which was produced using Pythia 6.4.
The $t\bar{t}$ cross section was taken from \cite{Bonciani:1998vc}, while the cross sections for the other backgrounds are
calculated from Madgraph. The subsequent parton shower and hadronization were
simulated by Pythia 6.4. We have used the CKKW matching scheme
\cite{Catani:2001cc} implemented in Madgraph.  The events are then passed on to
PGS-4 \cite{PGS4} with  parameters chosen to mimic a generic
ATLAS/CMS type detector. All background sources considered, and their respective
cross sections are given in Table~\ref{tab:SMBCrossSections}.

The signal event samples, for gluino pair production and decay,  were
produced using Pythia 6.4 and have been passed through the same PGS-4
detector simulation. Appropriate k-factors \cite{Beenakker:1996ch}
were applied to the LO signal cross-section calculated by Pythia to
obtain the NLO cross-section.

Basic muon isolation was applied to all samples: If the summed $P_{T}$ in a
$\Delta R=0.4$ cone around the muon is greater than 5 GeV, or the ratio of the
$E_{T}$ in a $3\times3$ cell region of the calorimeter to the muon $P_{T}$ is
greater than 0.1125, the muon is merged with the nearest jet in $\Delta R$.

We have also imposed on both the signal and the background the
following selection cuts
\begin{enumerate}
\item ${\not}E_{T}$$\ge100$ GeV
\item $p_{T}\ge20$ GeV and pseudorapidity $|\eta|<2.5$ for all objects
\item At least $4$ jets with $p_{T}\ge50$ GeV
\end{enumerate}

\begin{table}
\begin{centering}

\begin{tabular}{c}
 \textbf{Standard Model Background} \\
\begin{centering}
\begin{tabular}{cccccc}
  & \emph{0b} & \emph{1b} & \emph{2b} & \emph{3b} & \emph{$\geq $4b} \\
 \emph{0L} & 1717.46 & 3069.29 & 2091.27 & 320.54 & 36.51 \\
 \emph{1L} & 783.34 & 1489.85 & 998.8 & 118.42 & 8.49 \\
 \emph{OS} & 41.89 & 61.82 & 34.06 & 4.46 & 0.01 \\
 \emph{SS} & 0.41 & 0.97 & 0.44 & 0.04 & - \\
 \emph{3L} & 0.1 & 0.54 & 0.24 & 0.06 & - \\
 \emph{$\geq $4L} & - & - & - & - & -
\end{tabular}
\end{centering}
\end{tabular}

\end{centering}

\caption{\label{tab:nev-bg} Number of Standard Model events with $n$
  ($n=0..4$) b-tagged
jets and $m$ ($m=0,1,OS,SS,3,4$) leptons for the combined SM background
considered. The following cuts were applied: MET $\ge$ 100 GeV, at least 4 
jets
with $p_T\ge 50$ GeV, all jet and lepton $p_T \ge$ 20 GeV. The results
are normalized to $500$ pb$^{-1}$.}
\end{table}

\begin{table}[h!]

\begin{centering}

\begin{tabular}{ccc}
 \textbf{Model A} &  \textbf{Model C} & \textbf{Model D} \\

\begin{tabular}{cccc}
  & \emph{2b} & \emph{3b} & \emph{$\geq $4b} \\
 \emph{1L} &  & 166.2 & 70.9 \\
 \emph{OS} & 27.6 & 19.3 & 7.3 \\
 \emph{SS} & 12.7 & 9.4 & 3.1 \\
 \emph{3L} & 3.1 & 2.2 & 0.8
\end{tabular}
 &
\begin{tabular}{cccc}
  & \emph{2b} & \emph{3b} & \emph{$\geq $4b} \\
 \emph{1L} &  & 106.5 & 44.2 \\
 \emph{OS} & 13.3 & 10. & 3.9 \\
 \emph{SS} & 4.1 & 2.8 & 1. \\
 \emph{3L} & 1.1 & 0.6 & 0.2
\end{tabular}
 &
\begin{tabular}{cccc}
  & \emph{2b} & \emph{3b} & \emph{$\geq $4b} \\
 \emph{1L} &  & 98. & 37.8 \\
 \emph{OS} & 5.6 & 3.8 & 1.5 \\
 \emph{SS} & 2.9 & 2.2 & 0.8 \\
 \emph{3L} & 0.2 & 0.1 & 0.1
\end{tabular}

\end{tabular}

\end{centering}
\caption{\label{tab:nev-sig} Number of signal events passing the
selection cuts and containing $n$  b-tagged
jets and $m$ leptons.
The selection cuts applied were: MET $\ge$ 100 GeV, at least 4 jets with 
$p_T
\ge50$ GeV, all jet and lepton $p_T \ge$ 20 GeV. The results are normalized to
$500\; pb^{-1}$.}

\end{table}

Table \ref{tab:nev-bg} shows the expected number of events from the SM
background. We have classified them according to the number of
$b$-tagged jets and isolated leptons in the event. Same sign (SS) and
opposite sign (OS) di-leptons are separated as they have very different origins and sizes. We will
use the possible excess in these channels to assess the discovery potential.
The results are normalized to $500\; pb^{-1}$ of integrated luminosity.
Crossed out entries indicate no background events passing the
signature and selection cuts.

\begin{table}

\begin{centering}

\begin{tabular}{ccc}
 \textbf{Model A} & \textbf{Model C}  & \textbf{Model D} \\

\begin{tabular}{cccc}
  & \emph{2b} & \emph{3b} & \emph{$\geq $4b} \\
 \emph{1L} &  & 15.3 & 24.3 \\
 \emph{OS} & 4.73 & 9.12 & 87.0 \\
 \emph{SS} & 19.2 & 49.4 & - \\
 \emph{3L} & 6.44 & 9.26 & -
\end{tabular}
 &

\begin{tabular}{cccc}
  & \emph{2b} & \emph{3b} & \emph{$\geq $4b} \\
 \emph{1L} &  & 9.79 & 15.2 \\
 \emph{OS} & 2.28 & 4.73 & 47.0 \\
 \emph{SS} & 6.10 & 14.5 & - \\
 \emph{3L} & 2.35 & 2.63 & -
\end{tabular}
 &
\begin{tabular}{cccc}
  & \emph{2b} & \emph{3b} & \emph{$\geq $4b} \\
 \emph{1L} &  & 9.00 & 13.0 \\
 \emph{OS} & 0.957 & 1.79 & 18.3 \\
 \emph{SS} & 4.31 & 11.3 & - \\
 \emph{3L} & 0.418 & 0.318 & -
\end{tabular}

\end{tabular}

\end{centering}

\caption{\label{tab:signif} Signal significance $S/\sqrt{B}$, computed for
the results in tables \ref{tab:nev-sig} and \ref{tab:nev-bg}. The
crossed out entries indicate
no background events passing the signature and selection cuts. }
\end{table}

Table \ref{tab:nev-sig}  shows the expected number of
signal events with $n$ $b$-tagged jets and $m$ isolated
leptons (leptons $=e^{\pm},\mu^{\pm}$). 
Model A, which is predominantly a four top signal, 
has significantly more muti-lepton
and $b$-jet events passing selection cuts than Model C and Model D, which 
have fewer four top events. Model C is a stronger signal than Model D,
 which has very few four top events.

In Table~\ref{tab:signif}, we show the signal significance achievable with 500
$pb^{-1}$ integrated luminosity. 
By requesting $\ge3$ $b$-tagged jets it 
is possible to observe signal significance $S/\sqrt{B}\ge$3 for events with
multiple leptons, an excess consistent with multi-top production. In the single
lepton channels, a more detailed study of the background would be required to
carefully calculate the expected significance, since there are likely to be
significant background contributions from QCD processes with a faked lepton.  The
same-sign dilepton channel is probably the best channel for discovery, a finding
that is consistent with results in \cite{samesign_dilepton}.
It can also be observed that already with 100 pb$^{-1}$ integrated
luminosity a 4 top signal (Model A) may be established in the same
sign dilepton, 3 $b$-jet channel.

During the early period of LHC data taking, missing energy may not be well understood since it requires a 'global' understanding of the
ATLAS/CMS detectors. Therefore missing energy 
should not necessarily be taken as a reliable tool to discover new physics at low luminosities. If we 
do not include the missing energy cut in our analysis, then QCD backgrounds, particularly $b\bar{b}$ production, becomes a significant background to the multi-top 
signal. Requiring four hard jets, as we have done here, does reduce the QCD backgrounds since hard jets are less likely to produce isolated leptons 
\cite{Baer:2008kc}. However, even though the 2-lepton background from QCD might still be significant, 
the 3-lepton QCD background will unlikely be more significant than the 4-top signal 
\cite{Baer:2008kc,Sullivan:2008ki}. Thus, it seems reasonable that, without using missing energy,
discovery could still be possible without missing energy at integrated luminosities greater than or equal to $500 \mbox{ pb}^{-1}$.

\section{\label{sec:reconstruction}Direct Reconstruction}

Once evidence is obtained for an excess beyond the Standard Model in events with
multiple $b$-jets and leptons, it is natural to assume that the signal involves
production of multiple top quarks. In order to provide concrete evidence for
this, $W$ bosons and top quarks in the signal should be reconstructed.

In the model under consideration, where each signal event has four top quarks,
the main sources of difficulty in reconstruction are low statistics, large
combinatorial background, and poor object reconstruction due to the
extremely complex event topology. Every $tttt$
event has four $W$ bosons, each of which gives two jets if it decays
hadronically. Furthermore, every top decay itself gives a $b$-jet.
Therefore the expected number of hard jets arising in a decay with $k$
$W$ bosons decaying hadronically is $4+2k.$ On top of this there are also
jets arising from initial/final state radiation (ISR/FSR).
For comparison, $t\bar{t}$ all hadronic decays are expected to have
$6$ hard jets on average before ISR/FSR effects are included.

We expect that due to the large number of particles in the final state,
the chance of reconstructible objects (such as jets) overlapping is
large, so that the detector will
frequently be unable to accurately reconstruct isolated objects.
If two partons are very close in $\Delta R$ (roughly less than twice
the diameter of the cone used in the jet algorithm), the jets coming
from the partons are likely to be recombined
into a single jet.  Similarly, if a lepton is very close to a jet, it
will likely not pass the isolation requirement.  We utilize the
standard cone jet reconstruction algorithm implemented
in the PGS simulator, however we reduce the $\Delta R$ cone size to 0.4 in line
with the expected performance of the ATLAS detector.

\begin{figure}
\centering
\subfigure[]{
	\includegraphics[scale=0.35]{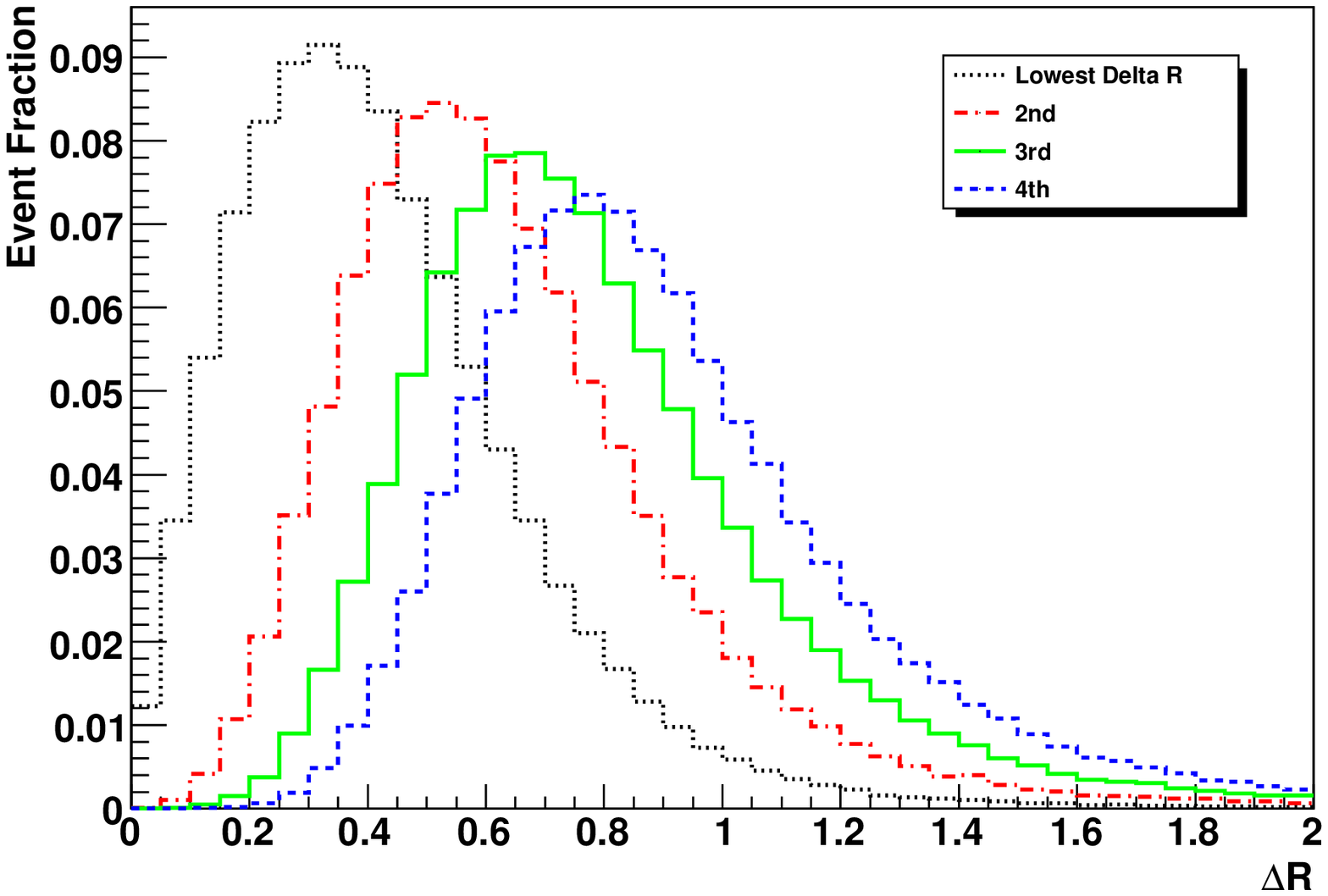}}
\hspace{0.1in}
\subfigure[]{
	\includegraphics[scale=0.35]{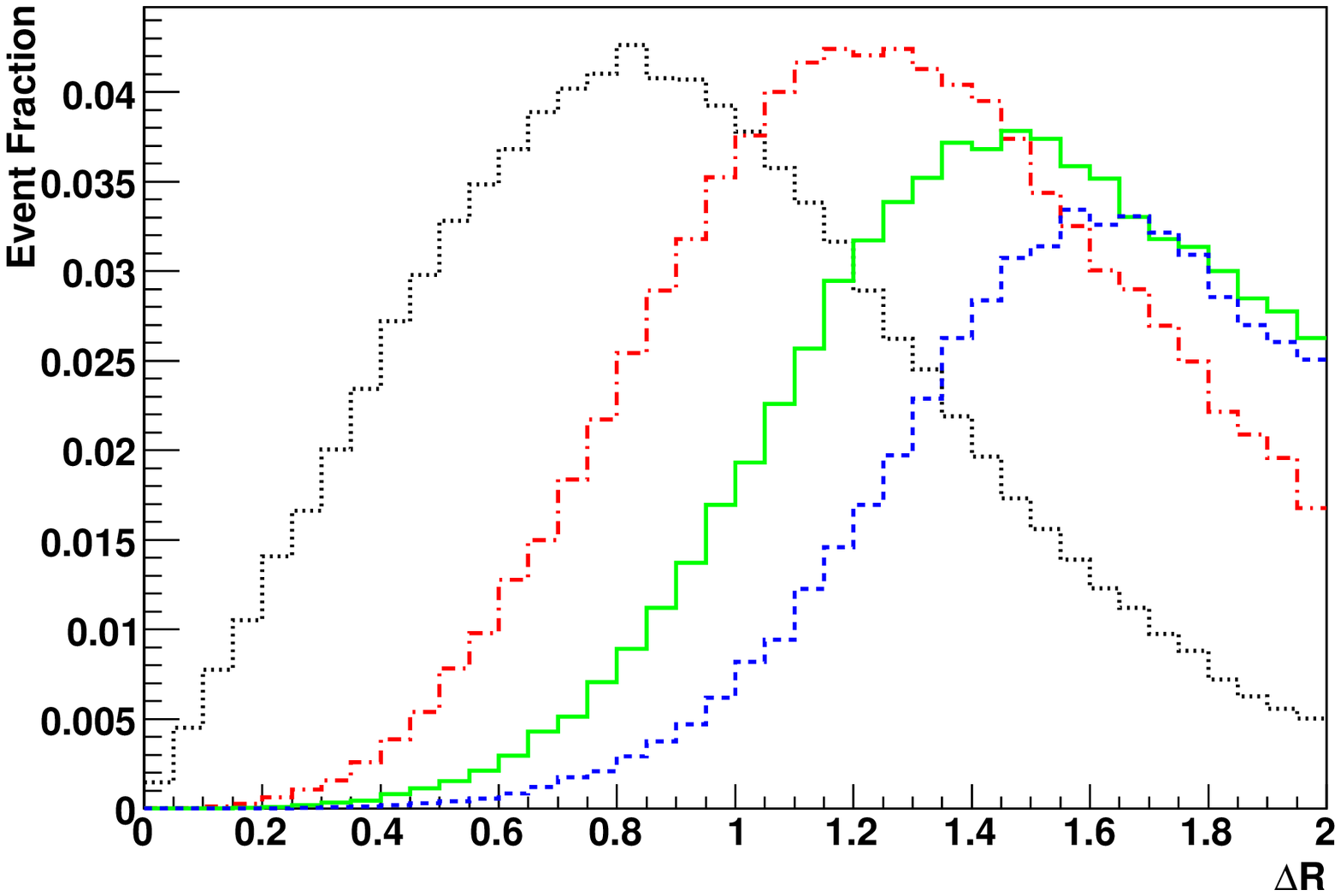}}	
{\caption{\label{overlaps}  Distributions of the lowest four values of
    separations, $\Delta
R$, between reconstructible partons (which here we take as any first/second
generation quark or b-quark) obtained using the Pythia parton-level (truth)
information for benchmark model A.  For each event in the simulation, $\Delta R$
was computed for all pairs of reconstructible objects, ranked, and the lowest
four values binned into histograms. The resulting distributions showing the
lowest, 2nd, 3rd, and 4th lowest $\Delta R$ values encountered are given for (a)
four-top events of model A, and also (b) hadronic $t\bar{t}$ decays.  All
distributions are normalized to unity.  The partons present in four-top events
are significantly closer to one another relative to those in SM $t\bar{t}$
events, increasing the liklihood of overlap, as well as a lower reconstruction
efficiency.}} \end{figure}

Figure \ref{overlaps} shows the distributions of the lowest, 2nd lowest,
3rd lowest, and 4th lowest $\Delta R$ values between reconstructible partons in
four top events from benchmark model A.  For comparison, the same information
is also shown for Standard Model $t\bar{t}$ events. Here we define a
reconstructible parton as any first/second generation quark or $b$-quark. It may
be seen in figure \ref{overlaps} (a) that in four top events there
is a large likelihood of three to four pairs of overlapping
objects, rendering reliable final state reconstruction difficult.

Direct reconstruction of top quarks in somewhat different decay chains
has been studied
in Ref.~\cite{Hisano:2002xq,Hisano:2003qu}. We expect a similar study
in our case will also yield at least some evidence that there are
indeed top quarks in the event. Rather then presenting a detailed
analysis here, we  focus here on a somewhat different
question. Clearly, in order to completely measure the branching ratios
into different final states involving different numbers of top and
bottom, it is not enough to just reconstruct a certain top quark. We
need to be able to reconstruct {\it all} of the top quarks in the
event with reasonable efficiency. However, our study already shows that
there is a significant overlap between different object in this type
of signal events. Including additional combinatorics, we expect
a very low efficiency for reconstructing multiple tops.

To gain some estimate of such efficiencies, we study how many tops we can
possibly reconstruct in a event.  To proceed, we define a 'top candidate' to be
the combination of two light-jets (ie non-$b$-tagged) and one $b$-tagged jet, where the
non-$b$-tagged dijet invariant mass satisfies $65 < m_{jj} < 95$ GeV, the
invariant mass of the b-jet with any lepton must satisfy $m_{bl} > 155$ GeV,
while the invariant mass with either of the two non-$b$-tagged jets must satisfy
$m_{bj} < 160$ GeV. Finally, the invariant mass of the final three-jet
combination must satisfy $125 < m_{jjb} < 225$ GeV.

\begin{figure}
\centering
\subfigure[]{
	\includegraphics[scale=0.35]{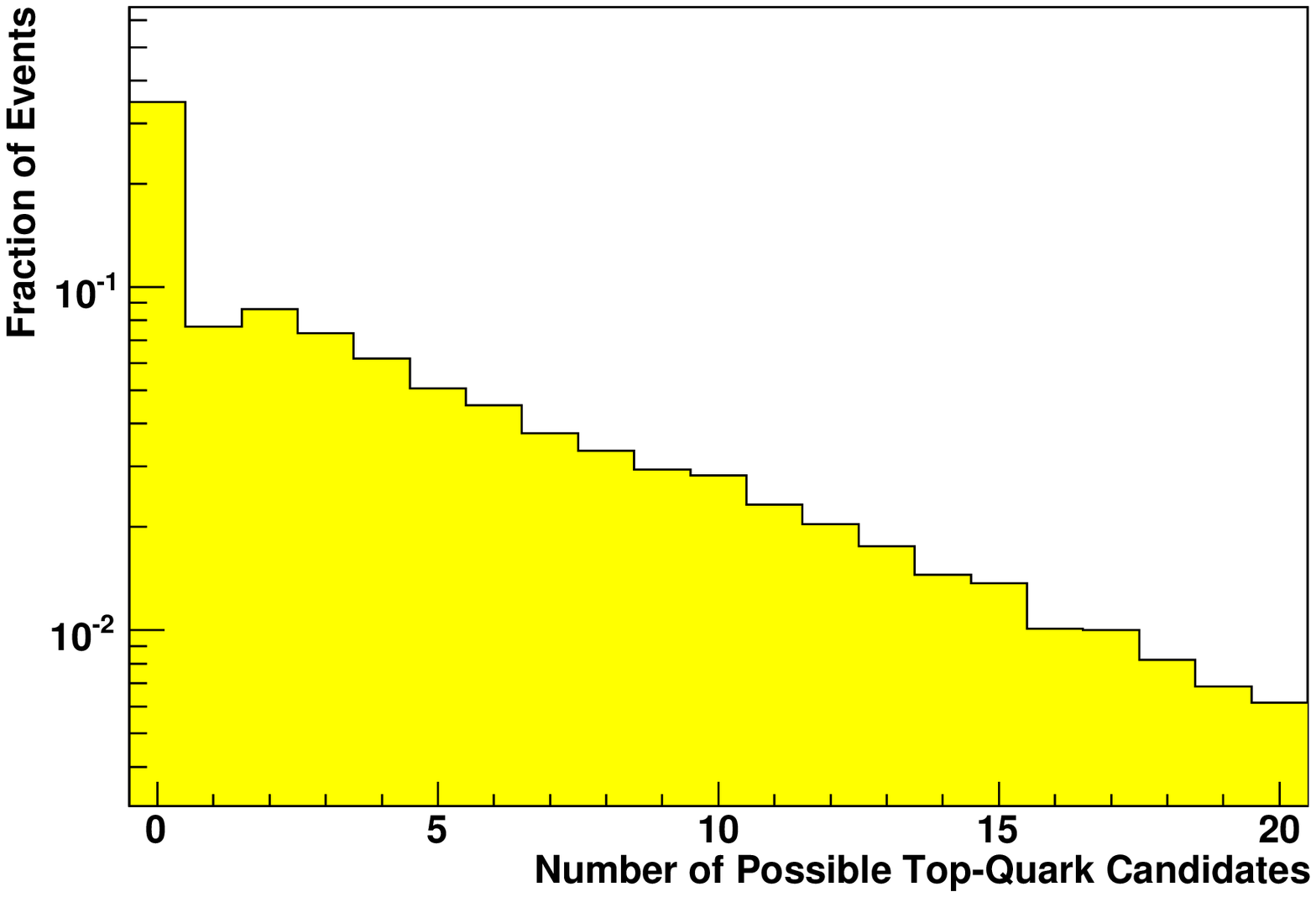}}
\hspace{0.1in}
\subfigure[]{
	\includegraphics[scale=0.35]{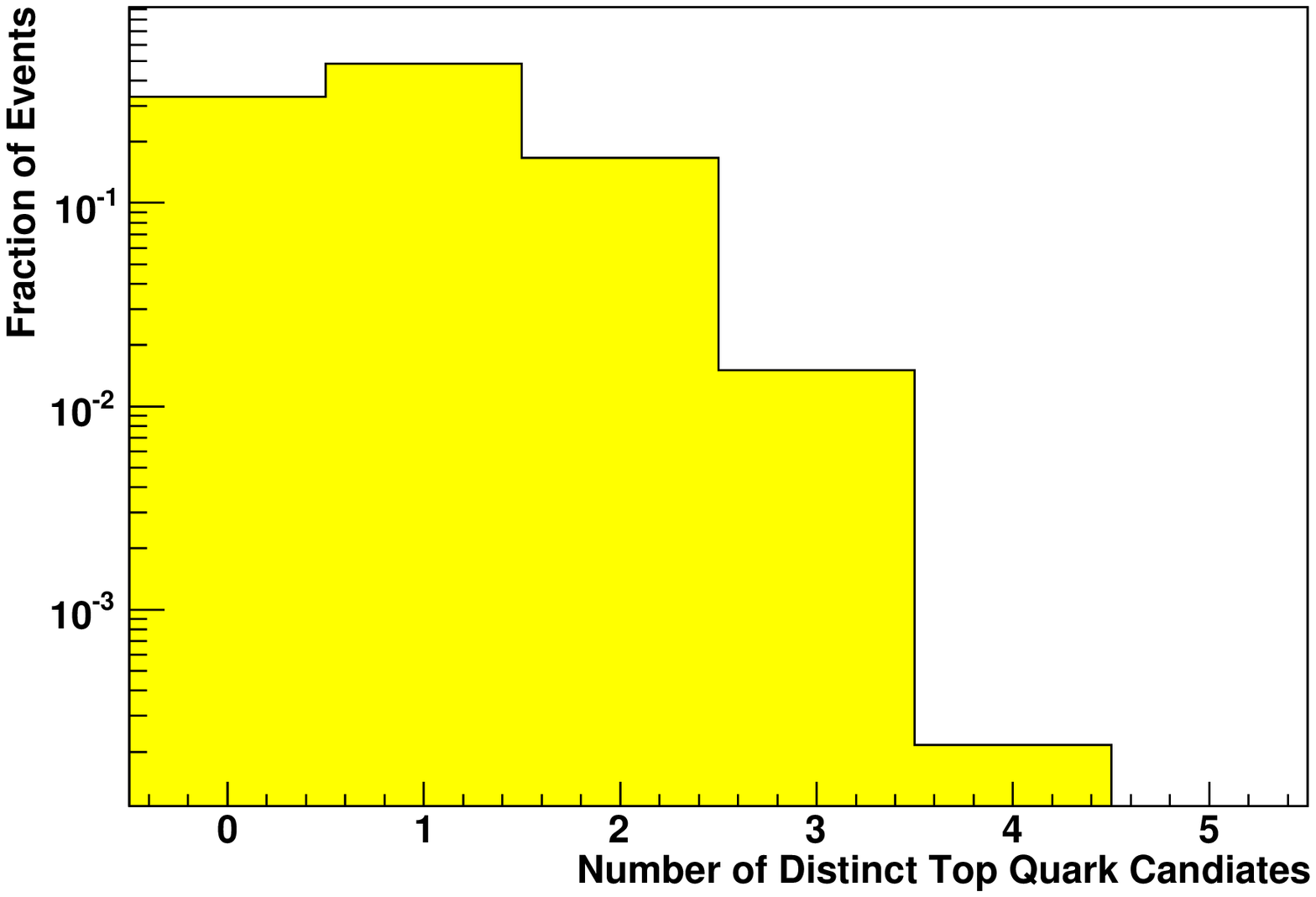}}	
{\caption{\label{fig:tc dist} As a demonstration of the combinatoric background,
in figure (a), we show the observed number of possible top 'candidates' obtained
for our benchmark model A. (a) We consider combinations with one $b$-jet and 
two light jets, where the invariant mass falls within the top mass window, and where
the combination satisfies a minimal set of selection criteria (see text). 
Due to the combinatorics, the
same $b$-quark can be combined with other partons to form multiple ``top
candidates''. In (b) we show  the resulting number of top candidates after
attempts are made to remove combinatorics by isolating distinct 3-jet
combinations.  For this basic study we only require that events have at least 4
$b$-tagged jets.  }} \end{figure}

Figure \ref{fig:tc dist} shows two distributions for the number of top
candidates observed in our benchmark model A.  The first figure, \ref{fig:tc
dist} (a), includes all possible three-quark combinations that satisfy the
requirements above.  The inherent combinatorical background due to the intense
hadronic activity in four-top events is clearly visible. We see that
the same $b$-quark can be combined with other partons to form several
``top candidates''.  There can be no more
than 4 top quarks present in the event, and the vast majority of 'candiate'
combinations are incorretly chosen.  Figure \ref{fig:tc dist} (b) shows the same
information except that here we isolate only the distinct jet combinations of
each event. Degeneracies that arise are removed by keeping track of the mass
difference $m_{jjb} - m_t$ for each jet triplet, and choosing the set of
triplets with the lowest average difference.  From the figure, it is clear that
this approach gives a significantly more reasonable result.  However, notice
that the number of reconstructed candidates drops dramatically as the top
multiplicity increases, rendering statistical analysis essentially impossible
without a large integrated luminosity.

The study we perform here is not an actual reconstruction of the top quark. A
true reconstruction involves positive identification of the top quark through
statistical determination of the top invariant mass. Instead, this study
is an estimate of how many objects obtained from recombining final
states can be consistent with a top quark. We expect this study,
though not completely precise, does capture the main effect of
combinatorics and object merging. We observe that the
efficiency for detecting one top quark as defined is approximately $48.5\%$,
for two quarks $\sim 16.7\%$, three quarks $\sim1.5\%$, and for four quarks
$\sim0.02\%$.

Further perfections of reconstruction techniques are certainly going
to improve these results and should be pursued. However, it is likely
that to extract statistically significant information based on a full
reconstruction will still require large integrated luminosity.

\section{Understanding multi-top final states}

\label{sec:fit}

As we have demonstrated in the previous sections, new physics signals
containing multiple tops can be discovered at the LHC. By direct reconstruction,
we expect to gather evidence that there are indeed top quarks in the signal.
However, there are many possible event topologies which can contribute
to our signal. In general, gluino can decay into third generation
quarks through $\tilde{g}\rightarrow t\bar{t}+\tilde{N}_{i}$,
$\tilde{g}\rightarrow b\bar{b}+\tilde{N}_{i}$,
and $\tilde{g}\rightarrow t\bar{b}+\tilde{C}_{i}^{-}$. Therefore,
final states coming from gluino pair production can involve from zero
to four top quarks, with relative amounts determined by gluino branching
ratios. As we have argued in the introduction, measuring such
branching ratios plays a central role in understanding
the properties of superpartners involved. For example, the relative
ratio of $\tilde{g}\rightarrow tt$ and $\tilde{g}\rightarrow bb$
could give us important information about the spectrum of the third
generation squarks. At the same time, significant
decay branching ratio of $\tilde{g}\rightarrow tb$ strongly
suggests that either Higgsino or Wino (or both) is lighter than the
gluino.

As demonstrated in the previous section, measuring the branching
ratios by directly reconstructing top quarks suffers from low
efficiencies. In this section, we will instead tackle this question
from a different approach. We assume that the new physics signal has
already been discovered
in a set of channels, particularly those with multiple leptons and
multiple bottom quarks (Table \ref{tab:nev-sig}). In addition, there
are a set of statistically significant
experimental observables defined on the set of signal events.
This general approach can in principle be applied in practice with
an arbitrary set of experimental observables based on availability.  Here,
we will only use a limited set of experimental variables
\begin{itemize}
\item 2 $b$-jets and either OS di-leptons, SS di-leptons, 3 leptons, 4
or more leptons
\item 3 $b$-jets and either 1 lepton, OS di-leptons, SS di-leptons, 3
leptons, 4 or more leptons
\item 4 or more $b$-jets and either 1 lepton, OS di-leptons, SS di-leptons,
3 leptons, 4 or more leptons
\end{itemize}

We will consider a general set of decay channels of the gluino which
can in principle contribute to the new physics signal. We will consider
as free parameters the relative branching ratios of those channels
whose values are determined by a fit to the above set of experimental
observables.
This approach can be viewed as a natural application of the method proposed
in Ref.~\cite{ArkaniHamed:2007fw}. The set of possible decay channels
are chosen as follows \[
\begin{array}{llll}
\tilde{g}\tilde{g}\rightarrow t\bar{t}\chi t\bar{t} &
\tilde{g}\tilde{g}\rightarrow t\bar{t}\chi t\bar{b}+c.c &
\tilde{g}\tilde{g}\rightarrow t\bar{t}\chi b\bar{b}\\
\tilde{g}\tilde{g}\rightarrow t\bar{b}\chi b\bar{b}+c.c. &
\tilde{g}\tilde{g}\rightarrow b\bar{b}\chi b\bar{b} &
\tilde{g}\tilde{g}\rightarrow t\bar{b}\chi\bar{t}b+c.c. &
\tilde{g}\tilde{g}\rightarrow t\bar{b}\chi t\bar{b}+c.c.
\end{array}\]
 where we continue to used the short hand notation of not explicitly
 displaying either the lightest neutralino, or the lightest
chargino. We have assumed that the lightest chargino and neutralino are
nearly degenerate so that transitions between them will not yield observable
decay products. $+c.c.$ indicates that we included the charge conjugated
event.

The number of events in the possible $\tilde{g}\tilde{g}$ decay channels
are given by \begin{eqnarray*}
n_{t\bar{t}t\bar{t}} & = & \sigma_{\tilde{g}\tilde{g}}\mathcal{L}Br(\tilde{g}\rightarrow t\bar{t})Br(\tilde{g}\rightarrow t\bar{t})\\
n_{t\bar{t}t\bar{b}} & = & \sigma_{\tilde{g}\tilde{g}}\mathcal{L}Br(\tilde{g}\rightarrow t\bar{t})Br(\tilde{g}\rightarrow t\bar{b})\\
\vdots & \vdots & \vdots\\
n_{qqqq} & = & \sigma_{\tilde{g}\tilde{g}}\mathcal{L}Br(\tilde{g}\rightarrow qq)Br(\tilde{g}\rightarrow qq).\end{eqnarray*}
These number can then be used to estimate the number of observed events
with a particular signature, $N_{{\rm obs}}^{{\rm sig}}$, which can
receive contributions from several channels listed above with a particular
fraction, $\epsilon_{{\rm channel}}^{{\rm sig}}$, depending on event
topology and experimental efficiencies. We have \begin{eqnarray*}
N_{{\rm obs}}^{0b0l} & = &
n_{t\bar{t}t\bar{t}}\epsilon_{t\bar{t}t\bar{t}}^{0b0l}+n_{t\bar{t}t\bar{b}}\epsilon_{t\bar{t}t\bar{b}}^{0b0l}+\ldots+n_{qqqq}\epsilon_{qqqq}^{0b0l}\\
N_{{\rm obs}}^{1b0l} & = &
n_{t\bar{t}t\bar{t}}\epsilon_{t\bar{t}t\bar{t}}^{1b0l}+n_{t\bar{t}t\bar{b}}\epsilon_{t\bar{t}t\bar{b}}^{1b0l}+\ldots+n_{qqqq}\epsilon_{qqqq}^{1b0l}\\
\vdots & \vdots & \vdots\\
N_{{\rm obs}}^{4b4l} & = & n_{t\bar{t}t\bar{t}}\epsilon_{t\bar{t}t\bar{t}}^{4b4l}+n_{t\bar{t}t\bar{b}}\epsilon_{t\bar{t}t\bar{b}}^{4b4l}+\ldots+n_{qqqq}\epsilon_{qqqq}^{4b4l}.\end{eqnarray*}
In the rest of this section, we will first obtain estimates of all
signal efficiencies $\epsilon_{{\rm channel}}^{{\rm sig}}$. Then,
we will perform a $\chi^{2}$-fit to determine a set of best fit values
of $\sqrt{\sigma_{\tilde{g}\tilde{g}}{\mathcal{L}}}\ Br_{{\rm
    channel}}$. Note
that in order to obtain the branching ratio from these counting
signatures, we will have to know the product
$\sigma_{\tilde{g}\tilde{g}}{\mathcal{L}}$.
Such information could be available independently from other measurements,
such as the gluino mass. We will show later in this section that our
method indeed gives us an estimate of this absolute rate. At this
moment, we note that we can already derive a lot of information
about the underlying model if we measure the ratio of branching
ratios, for which the dependence on
$\sigma_{\tilde{g}\tilde{g}}{\mathcal{L}}$ drops out.

The key step in this method of fitting for branching ratios is, of
course, to obtain an estimate the efficiencies,
 $\epsilon_{{\rm channel}}^{{\rm sig}}$.
Many factors enter in such an estimate. The existence of many objects,
in particular jets, in the event means that simple extrapolation from
single object efficiency is no longer reliable. In particular, there
is a significant chance now for the leptons from top decay to be close
to a jet and therefore fail the isolation cut. $b$-tagging can also
be affected. Numerical simulation using an appropriate model, a template, of the
underlying physics involved is therefore unavoidable.

In practice, certain assumptions about the underlying model must be
made in choosing a template. We have already chosen the set of channels
to include, based on information gained from the type of study in
the previous section. In addition, we have to choose the spectrum, the gluino mass and LSP mass,
to be used in the template. To begin with, we will first choose a template
model using the  {\it actual} gluino mass and LSP mass as the underlying
model. We will fit the underlying branching ratios by using
efficiencies obtained by simulating this template and demonstrate such
a fit does give us accurate information of the underlying model.  We will then
simulate a set of different templates with different mass
hypotheses. We will show that for the variation of mass hypotheses we have
studied, the difference induced for the fit does not significantly
affect our conclusion about the underlying models. Of course, such a
result still leaves the question of whether the range of variation of
mass parameters we have used is too optimistic. Later in this section,
we will show that indeed it is reasonable to expect we can get
estimates of mass scales from other experimental observables so that
we can choose our mass hypothesis with an error within this range. For
the continuity of our presentation, we will present our result for
the fits first. Then we will come back to address the question of
choosing the mass hypothesis in detail.

\begin{figure}
\includegraphics[scale=0.33]{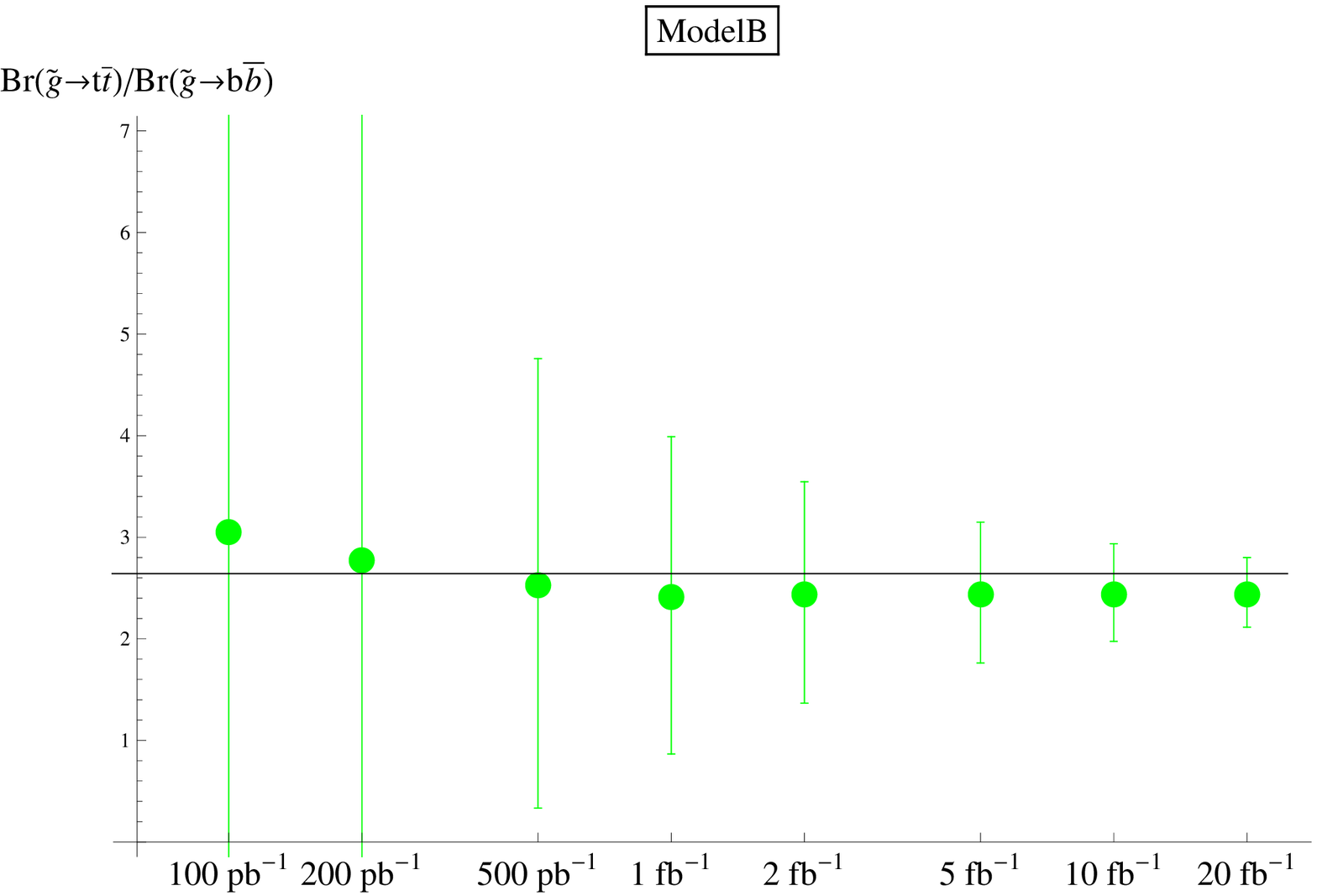}
\includegraphics[scale=0.33]{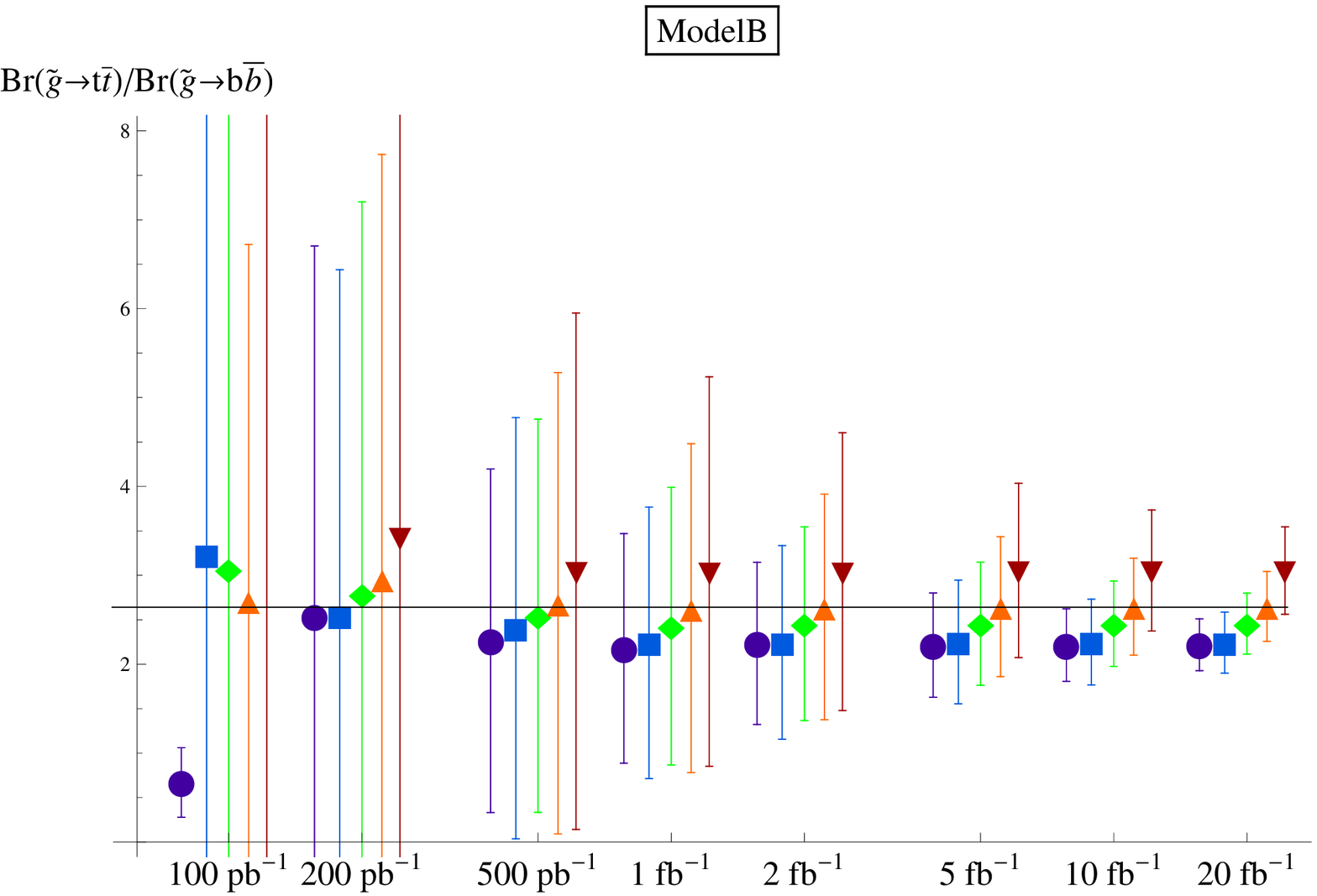}\includegraphics[scale=0.3]{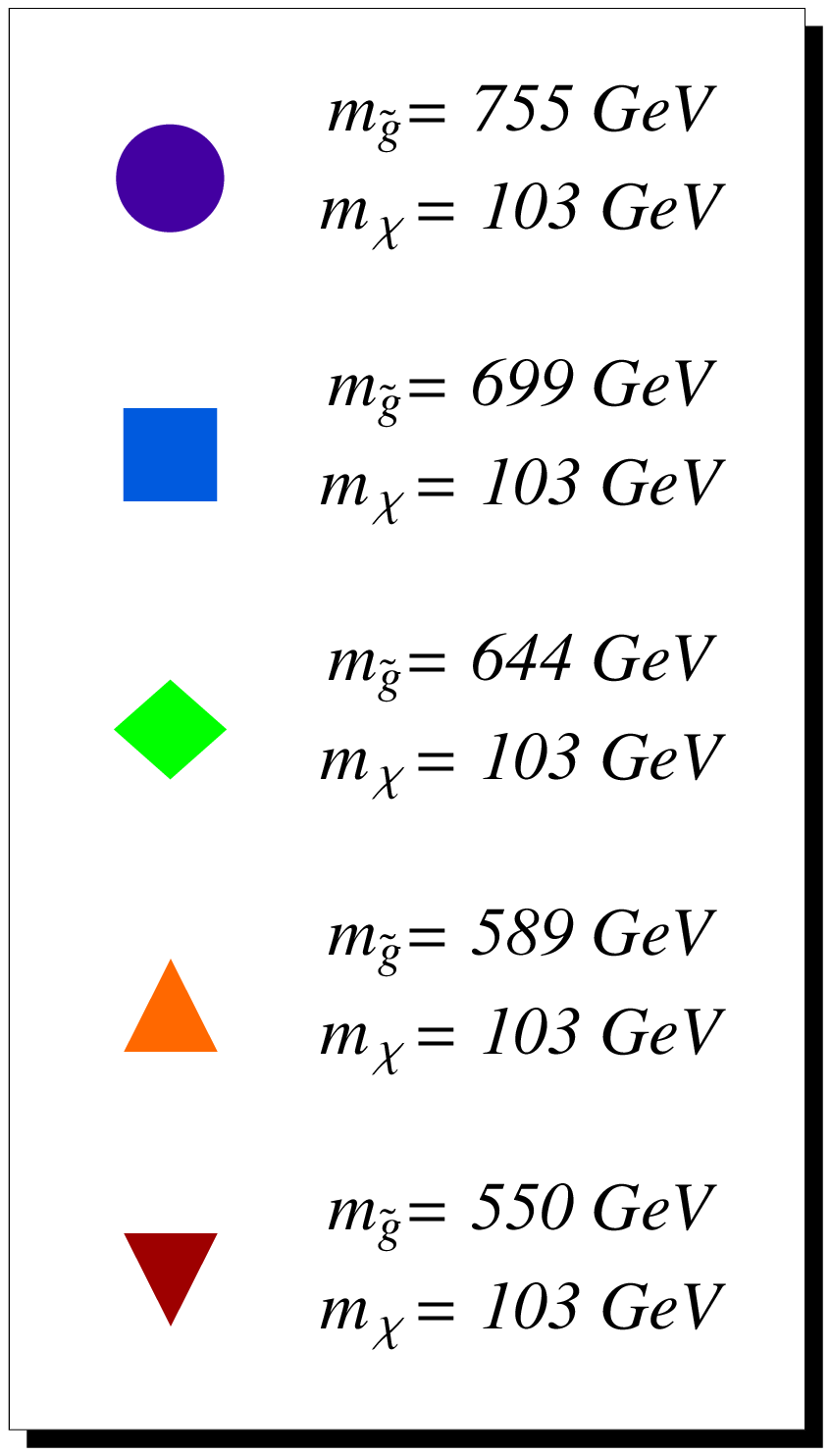}
\caption{Results for fits of ratio Br$(\tilde{g}\rightarrow t
  \bar{t})/$Br$(\tilde{g}\rightarrow b \bar{b})$  for benchmark
  Model-B. The left panel shows
  the result with the correct mass hypothesis. In the right panel,
the efficiencies are calculated for five different mass templates
show in the legend. The solid horizontal line gives the actual values
of the branching ratios. Errors are 1$\sigma$ and include errors
for subtracting off the $t\bar{t}$ background. \label{fig:fit-model-B}}
\end{figure}
Our fit to Br$(\tilde{g}\rightarrow t
\bar{t})/$Br$(\tilde{g}\rightarrow b \bar{b})$ for benchmark model-B
using the correct mass hypothesis
is shown in the left panel of Fig.~\ref{fig:fit-model-B}. For this and
all the other fits presented in this section, we
have required that for a channel to be included in our fits, there
must be at least 1 signal event, and the significance over the
\textbf{$t\bar{t}$ }background must be greater than 3.

We pause here to briefly describe how error bars are calculated in
these and the other fits presented in this section.
Using large statistics for the template model,
the statistical errors for calculating the efficiencies and for determining
the number of events for a given signature are assumed to be much
smaller that the expected Gaussian errors from the LHC data. The only
errors we include here are statistical errors from the minimization
procedure. Therefore, the $1\sigma$
error for a given branching ratio is given by the change in the
branching ratio required to shift the $\chi^{2}$ one unit from its
value at the minimum
\[
\delta_{Br(tt)}=\left.\left(\frac{1}{2}\frac{\partial^{2}\chi^{2}}{\partial
  Br(tt)^{2}}\right)^{-1/2}\right|_{\rm minimum},\;\;\delta_{Br(bb)}=\left.\left(\frac{1}{2}\frac{\partial^{2}\chi^{2}}{\partial
  Br(bb)^{2}}\right)^{-1/2}\right|_{\rm minimum}, \mbox{\ etc.}\]

From the left Fig.~\ref{fig:fit-model-B}, we see that using the
correct mass hypothesis, we will be able to measure the ratio
Br$(\tilde{g}\rightarrow t \bar{t})/$Br$(\tilde{g}\rightarrow b
\bar{b})$   with good accuracy for an integrated luminosity of $\sim
5-10$ fb$^{-1}$. In particular, we will be able to verify that in Model
B, the gluino decay is dominated by  $\tilde{g}\rightarrow t \bar{t}$
with a smaller but non-vanishing branching ratio for $\tilde{g}\rightarrow b
\bar{b}$.

Next, we want to assess the effect of changing our assumptions of
underlying spectrum. We will assume that although the mass spectrum
cannot be precisely measured, some crude estimates can still be made
based on kinematical variables such as $M_{{\rm eff}}$ and rate. We will
provide justifications for this assumption later in this section. Therefore,
we will consider cases where the gluino mass only deviates from
the underlying benchmark model by about 100 GeV. In particular, we
use four additional sets of alternative templates and carry out the
same fit. The result in Model B is shown in the right panel of
Fig.~\ref{fig:fit-model-B}.  We see that using different mass
hypotheses does make a visible difference. However, we observe that
these differences are not big enough to dramatically affect the
information we will extract from our measurement of
Br$(\tilde{g}\rightarrow t \bar{t})/$Br$(\tilde{g}\rightarrow b
\bar{b})$. In the case of 3-body gluino decay under consideration,
this ratio is proportional to
$(m_{\tilde{b}}/m_{\tilde{t}})^4$. Therefore, we see that using
different mass hypotheses within this range will at most result in a
factor two error in the measurement of the ratio of the branching
ratios, will induce at most $\sim 20 \%$ shift in the inferred ratio
$m_{\tilde{b}}/m_{\tilde{t}}$.

\begin{figure}
\includegraphics[scale=0.33]{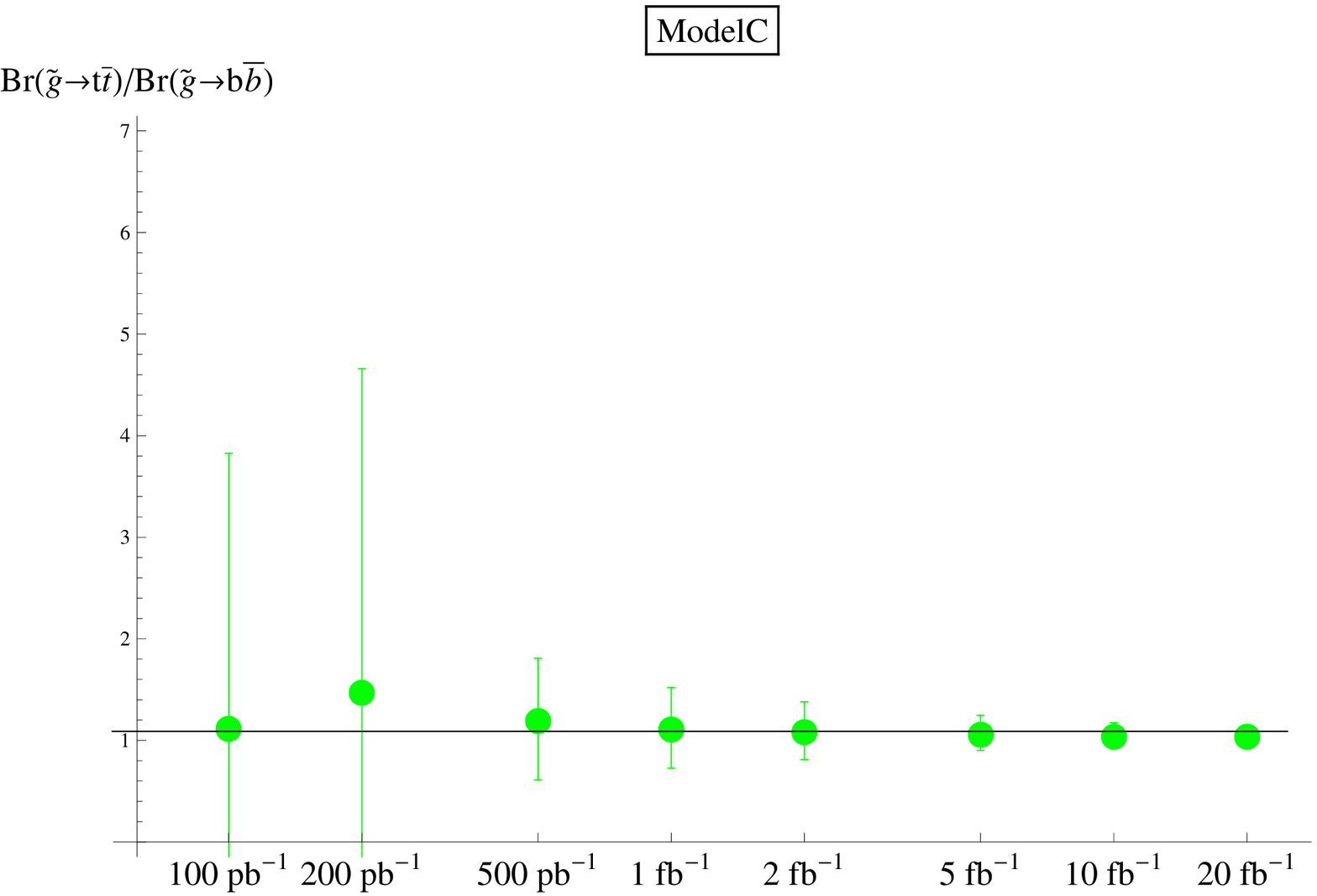}
\includegraphics[scale=0.33]{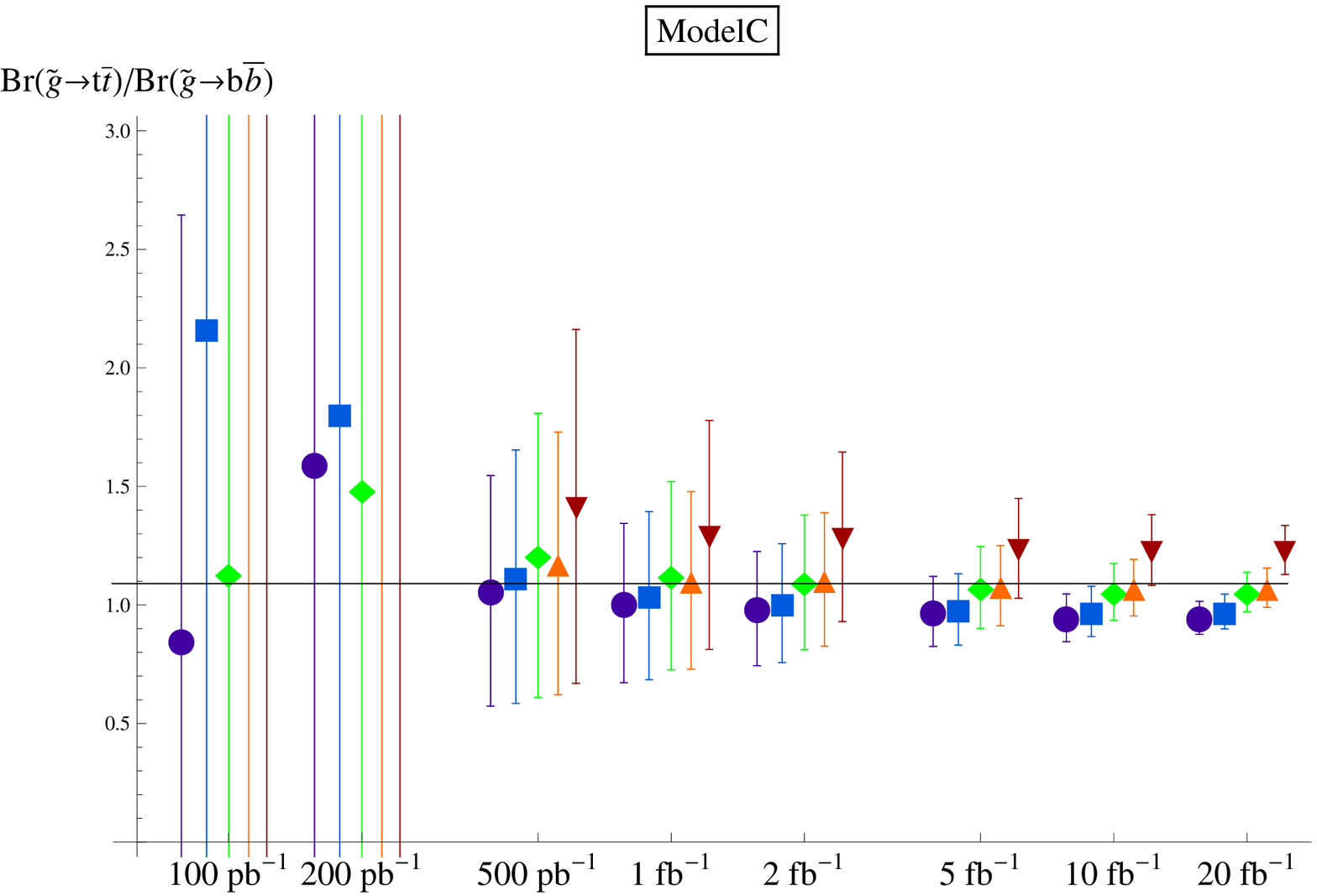}\includegraphics[scale=0.3]{Legend.eps}
\caption{Results for fits of ratio Br$(\tilde{g}\rightarrow t
  \bar{t})/$Br$(\tilde{g}\rightarrow b \bar{b})$  for benchmark
  Model-C. The left panel shows
  the result with the correct mass hypothesis. In the right panel,
the efficiencies are calculated for five different mass templates
shown in the legend. The solid horizontal line gives the actual values
of the branching ratios. Errors are 1$\sigma$ and include errors
for subtracting off the $t\bar{t}$ background. \label{fig:fit-model-C}}
\end{figure}
Notice that incorrect assumptions about the underlying spectrum do not
lead to significant effects in the fitting when considering at least
$1\mbox{ fb}^{-1}$ integrated luminosity of data. In all cases, we
can still extract a good estimate of the squark hierarchy and the
nature of the LSP.

Similar studies are performed for benchmark Model C. The result
for Model C is shown in Fig.~\ref{fig:fit-model-C}. Similar accuracies
of the measurements are obtained with same integrated luminosity. In
particular, we observe that we should
be able to distinguish Model B and C from these results alone with
about $5$ fb$^{-1}$ of integrated luminosity.

As a final note, we add that the even for the correct mass template
the ratio $Br(\tilde{g}\rightarrow t\bar{t})/Br(\tilde{g}\rightarrow
b\bar{b})$ is slightly underestimated in all cases. This is because
the data actually contains a small percentage of events in which the
gluino decayed to first and second generation quarks. As those decays
looks most similar the decay into two b-jets, the fit tends to
slightly overestimate $Br(\tilde{g}\rightarrow b\bar{b})$, and
therefore underestimating $Br(\tilde{g}\rightarrow
t\bar{t})/Br(\tilde{g}\rightarrow b\bar{b})$.

\begin{figure}[h!]
\includegraphics[scale=0.33]{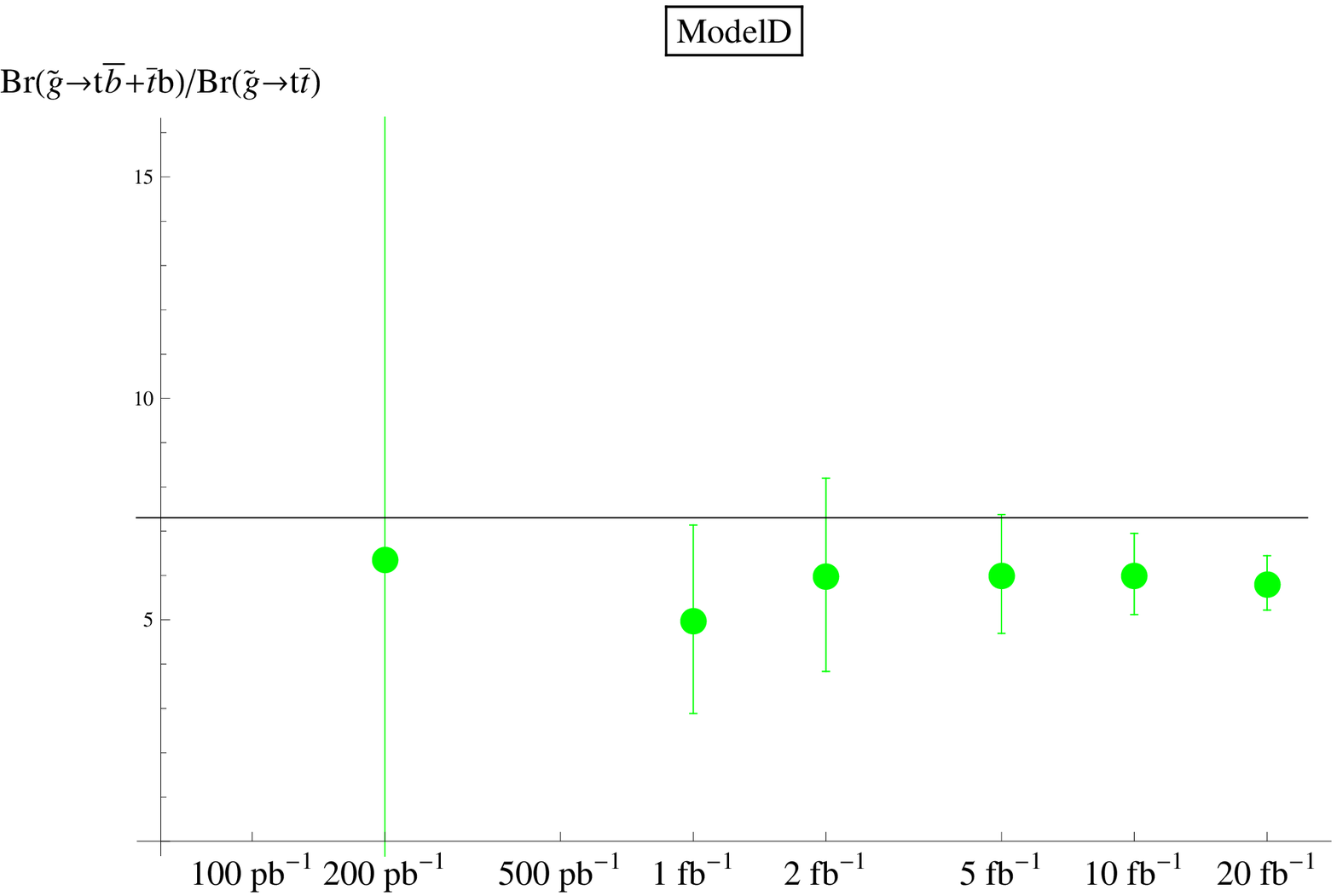}
\includegraphics[scale=0.33]{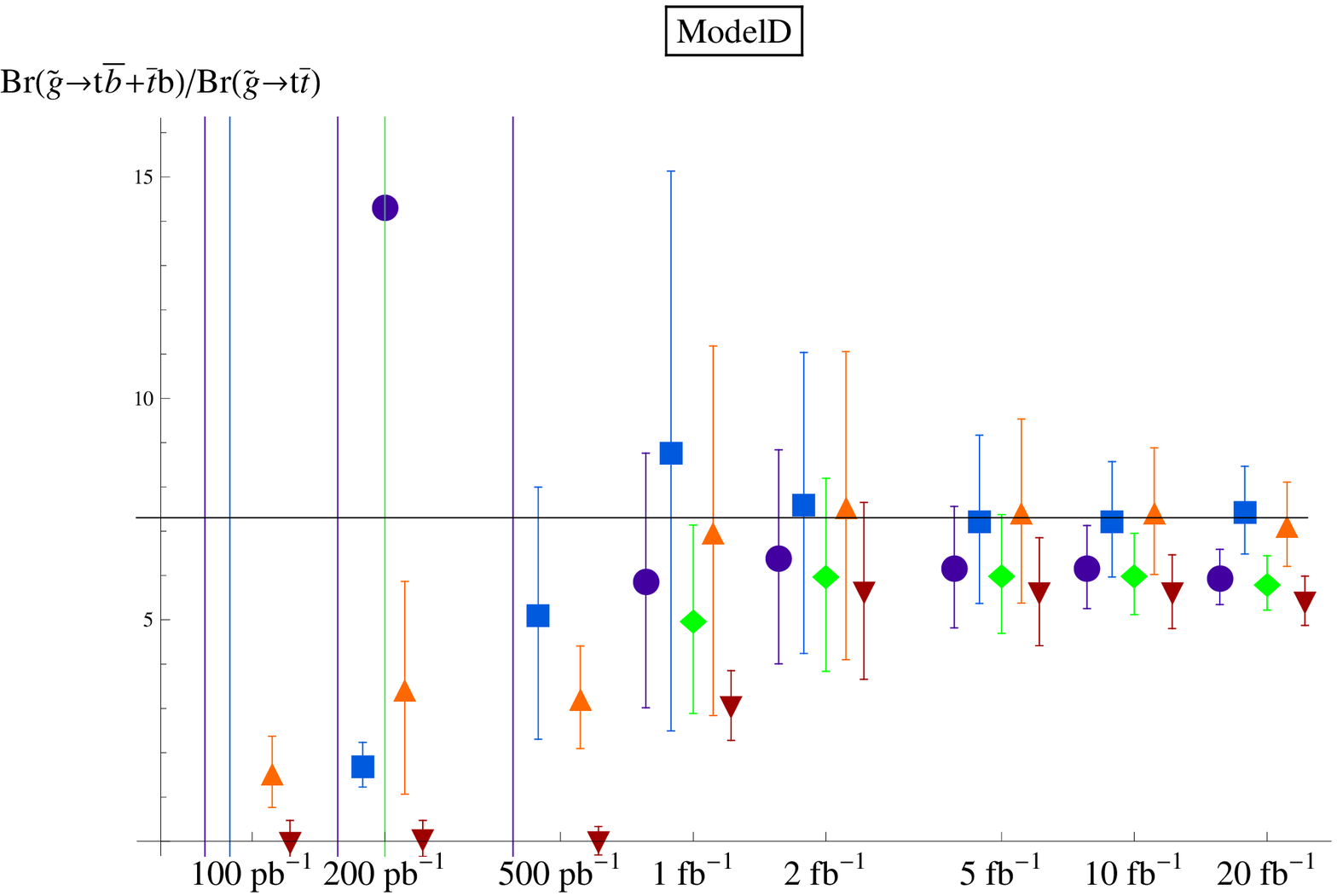}\includegraphics[scale=0.3]{Legend.eps}
\caption{Results for fits of ratio Br$(\tilde{g}\rightarrow t
  \bar{b}+\bar{t} b)/$Br$(\tilde{g}\rightarrow t \bar{t})$  for benchmark
  Model-D. The left panel shows
  the result with correct mass hypothesis. In the right panel,
the efficiencies are calculated for five different mass templates
show in the legend. The solid horizontal line gives the actual values
of the branching ratios. Errors are 1$\sigma$ and include errors
for subtracting off the $t\bar{t}$ background. \label{fig:fit-model-D}}
\end{figure}
The study of gluino decay for benchmark model D is presented in
Fig.~\ref{fig:fit-model-D}. In this case, we are interested in ratio
Br$(\tilde{g}\rightarrow t \bar{b}+\bar{t}
b)/$Br$(\tilde{g}\rightarrow t \bar{t})$. For a wino-LSP model like
Model D, we expect this ratio to be large, which can indeed be
experimentally verified with moderate luminosity, as shown in the figure.

We finally consider how well we can chose our mass
hypotheses based on available experimental data. This is certainly an
important issue since significantly wrong mass hypotheses lead to
misleading results. To begin with, we study the dependence  of our
result on the mass hypothesis in more detail.

\begin{figure}[h!]
  \begin{center}
    \includegraphics[scale=0.8]{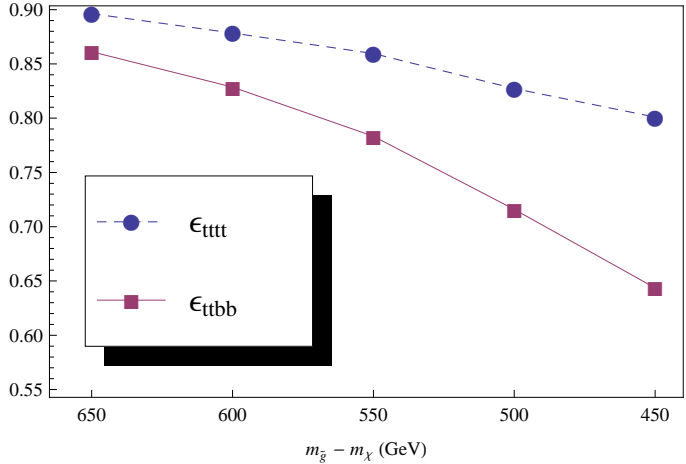}
  \end{center}
  \caption{Dependence of efficiencies on the mass gap $M_{\tilde{g}} -
    M_{\rm LSP}$ to pass the missing energy cut, MET $\ge$ 100 GeV. This dependance is the dominant effect for the variation of fits on the mass hypothesis.}
  \label{Eff}
\end{figure}

As the mass gap between the gluino and LSP is tightened, the events
will have a harder time satisfying the missing energy cut we
imposed. This more significantly affects events of the form
$\tilde{g}\rightarrow tt+{\not}E_{T}$ than $\tilde{g}\rightarrow
bb+{\not}E_{T}$, since the tops will use more of the gluinos' energy
than the bottoms. Thus a tighter mass gap used in our template
underestimates the ratio of efficiencies
$\epsilon_{t\bar{t}t\bar{t}}/\epsilon_{t\bar{t}b\bar{b}}$, see Figure
\ref{Eff}. The fitter then adjusts for this low efficiency by fitting
more $tttt$ events relative $ttbb$ to the signature counts
\[\frac{Br(\tilde{g}\tilde{g}\rightarrow t\bar{t} t\bar{t})}{Br(\tilde{g}\tilde{g}\rightarrow
  t\bar{t} b\bar{b})}\approx\frac{Br(\tilde{g}\rightarrow
  t\bar{t})^2}{2Br(\tilde{g}\rightarrow
  t\bar{t})Br(\tilde{g}\rightarrow
  b\bar{b})}\approx\frac{\epsilon_{t\bar{t}b\bar{b}}}{\epsilon_{t\bar{t}t\bar{t}}}.\]
Thus, in Models B and C as we increase(decrease) the gluino mass in our
templates we tend to underestimate(overestimate) the branching ratio
$Br(\tilde{g}\rightarrow t\bar{t})/Br(\tilde{g}\rightarrow
b\bar{b})$.

We see that the change in the mass gap between gluino and the LSP can
account for most of the variation in the fit result.
For example, extrapolating from our study, if our
assumption of gluino mass is off by more than 200 GeV, the result for
benchmark Model B will indeed look similar to that of Model
C. However, since we expect to have significant excess in multiple
channels, we observe that we should also have a significant amount
of information about the mass scales, and in particular the mass gap
$M_{\tilde{g}} - M_{\rm LSP}$ of the new physics
particles already at early stages of LHC running. For example, we would expect
simple transverse variables in channels with multiple leptons (and hence
lower Standard Model background) should already provide some
indication of the mass scales involved.

\begin{figure}
\begin{center}
\includegraphics[scale=0.4]{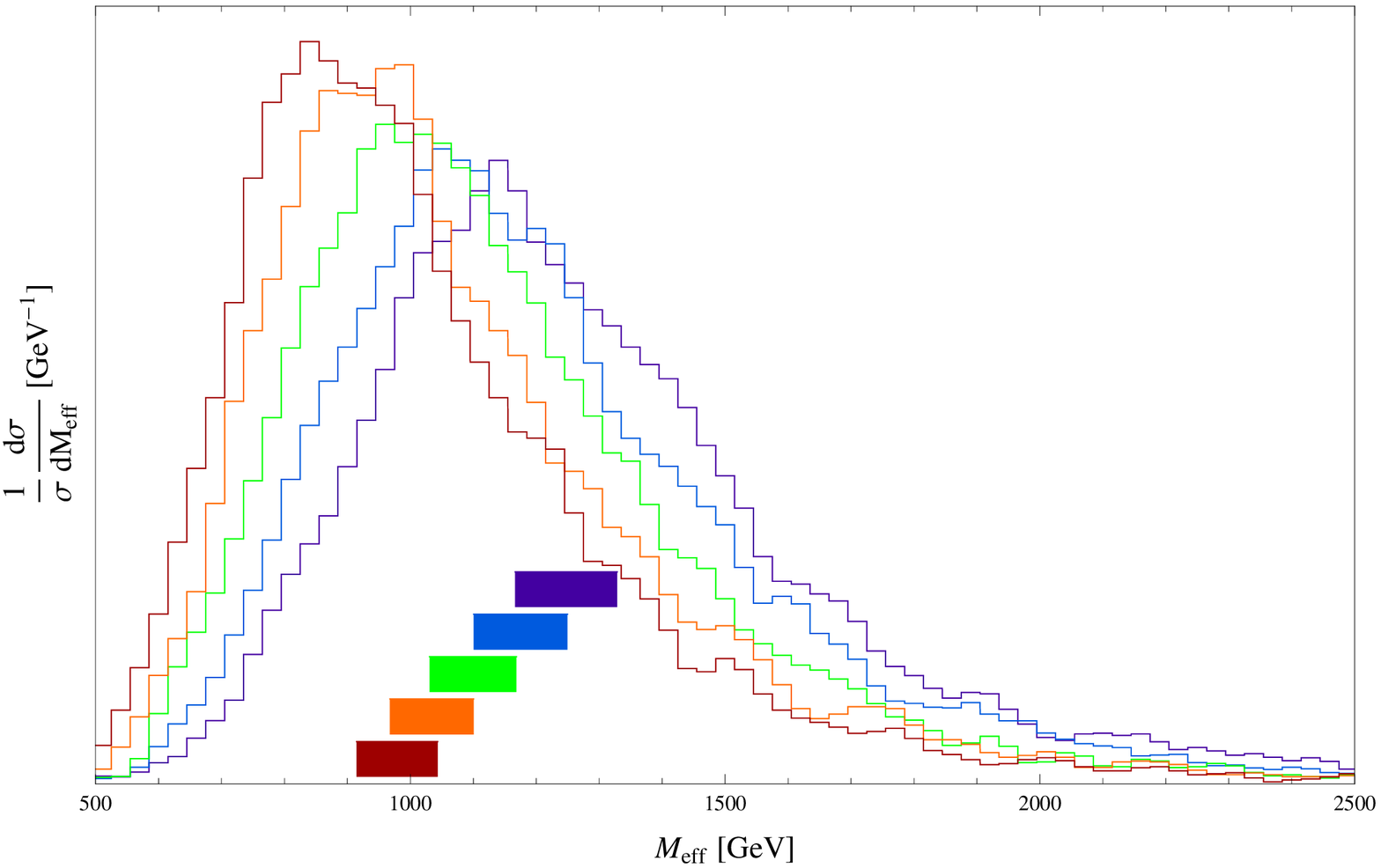}\includegraphics[scale=0.325]{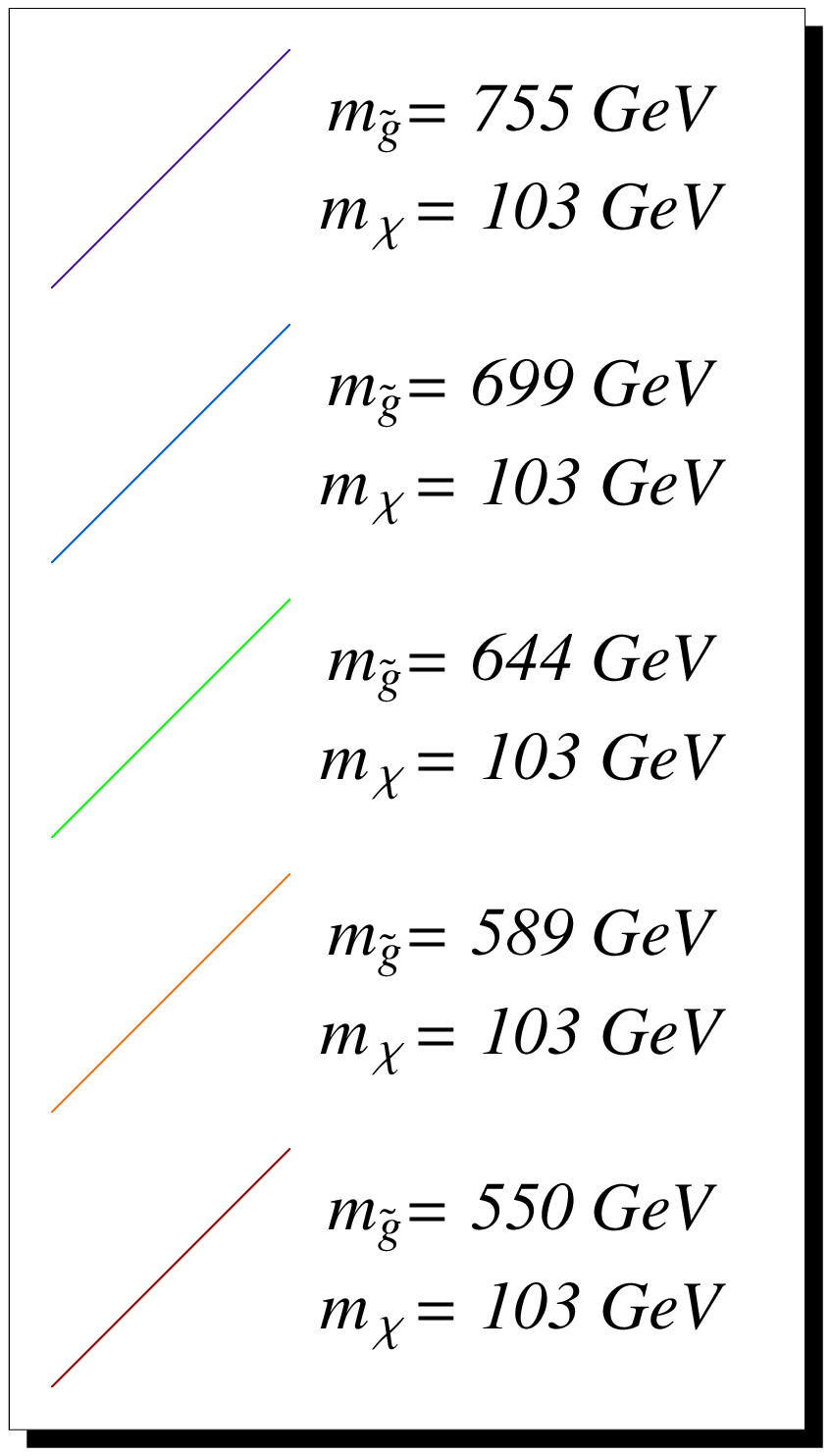}
\end{center}
\caption{\label{fig:EffMass-GluinoMass}Effective mass distribution, of
  $\tilde{g}\tilde{g}\rightarrow \bar{t}t\bar{t}$ events for the 5
  gluino masses used in fits. The bars at the bottom show the location
  of the inner $20\%$ quantile; the highest one is for the largest
  gluino mass, the second highest for second largest gluino mass,
  etc. The histograms are normalized so that the total area is
  unity. }
\end{figure}

\begin{figure}[h!]
\begin{center}
\includegraphics[scale=0.4]{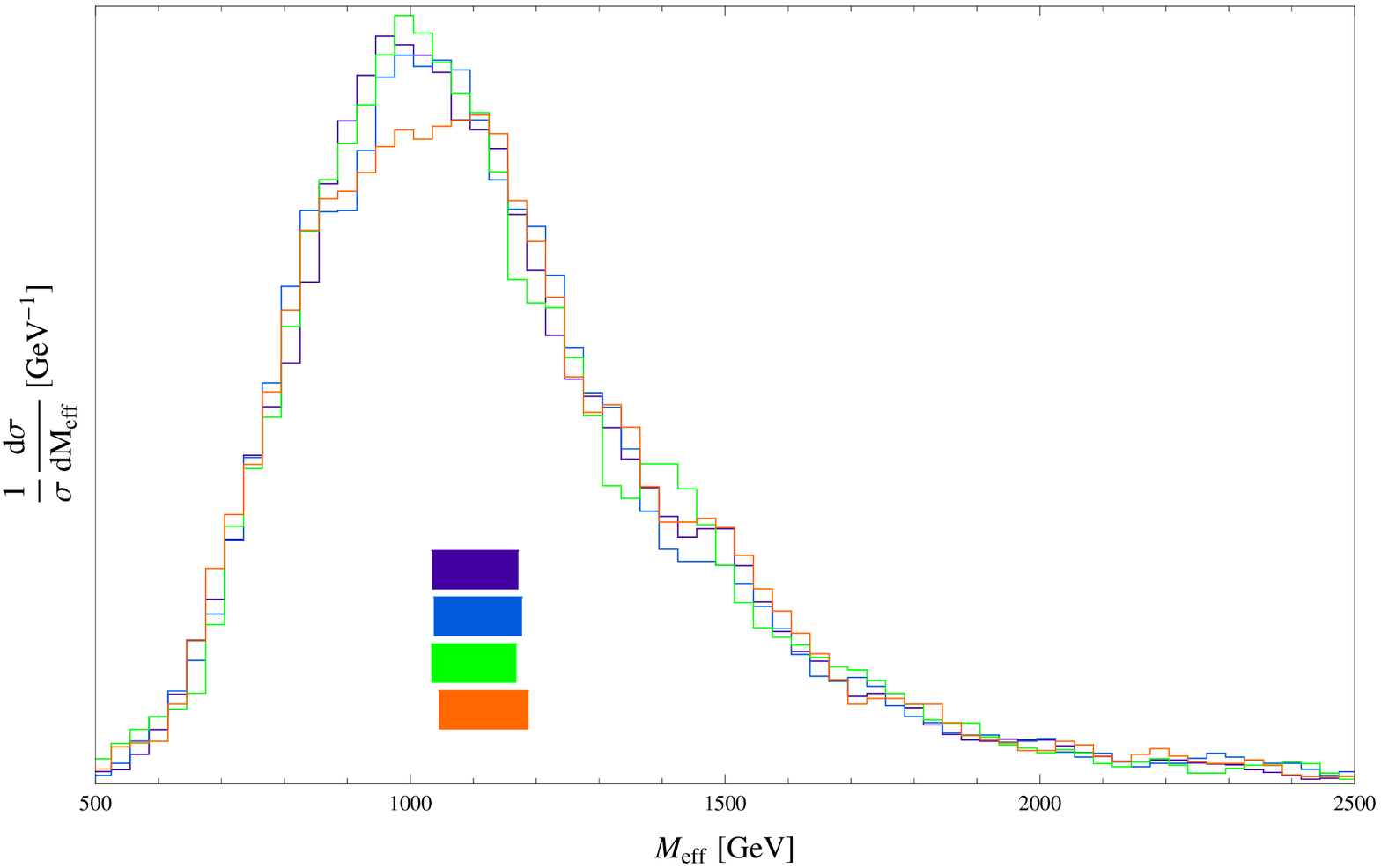}\includegraphics[scale=0.35]{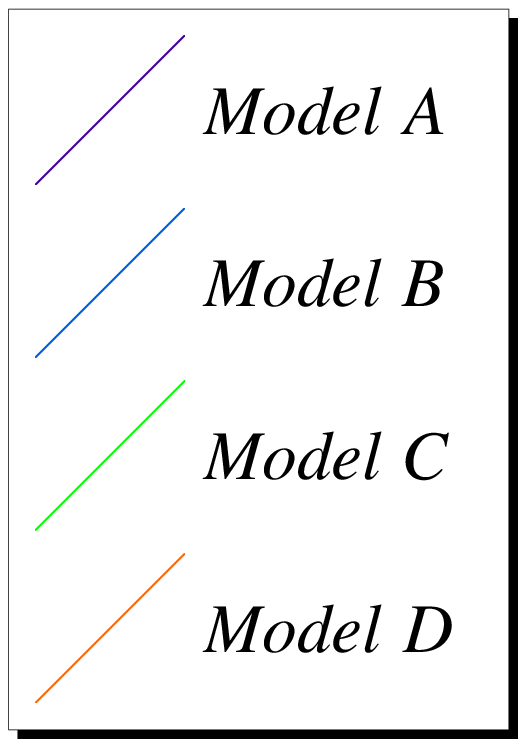}
\end{center}
\caption{\label{fig:EffMass-Model}Effective mass distribution, of
  Models-(A,B,C,D). The bars at the bottom show the location of the
  inner $20\%$ quantile. The
  histograms are normalized so that the total area is unity.} 
\end{figure}

Indeed, visible differences can be seen between different gluino masses
when we histogram the effective mass, $M_{\rm eff}$, defined as the
scalar sum of the transverse momentum for all objects in the an event
\[M_{eff} = \sum_{i} p_{T,i} + \displaystyle{\not}E_T  .\]
To demonstrate
this, we plotted the effective mass for $\tilde{g}\tilde{g}\rightarrow
t\bar{t}t\bar{t}$ for the 5 gluino masses used in fits above, in Figure
\ref{fig:EffMass-GluinoMass}. The histogram curves move to lower
energies (right to left) as the gluino mass is lowered. There is also
a noticeable change in the effective mass spread, which we quantify as
the location of middle $20\%$ quantile. The bars at the bottom show
the location of the inner $20\%$ quantile; the highest one is for the
largest gluino mass, the second highest for second largest gluino
mass, etc. The fact that the bars move right to left as the gluino
mass is lowered indicates that the median effective mass is
decreasing, while the fact that the bars are shrinking indicate the
effective mass has less variation as the gluino mass decreases.

This analysis was only carried out for $\tilde{g}\tilde{g}\rightarrow
t\bar{t}t\bar{t}$, so we still need to demonstrate that the effective mass is also
independent of the gluino decay. To do this we plotted the effective
mass for the four models considered in this paper, in Figure
\ref{fig:EffMass-Model}.  The bars at the bottom show the spread,
from top to bottom, for Model-A, Model-B, Model-C, and Model-D. As can
be seen, the four histogram curves are remarkably similar, and
have no noticeable differences in the median or spread.

\begin{figure}[h!]
\begin{center}
\includegraphics[scale=0.5]{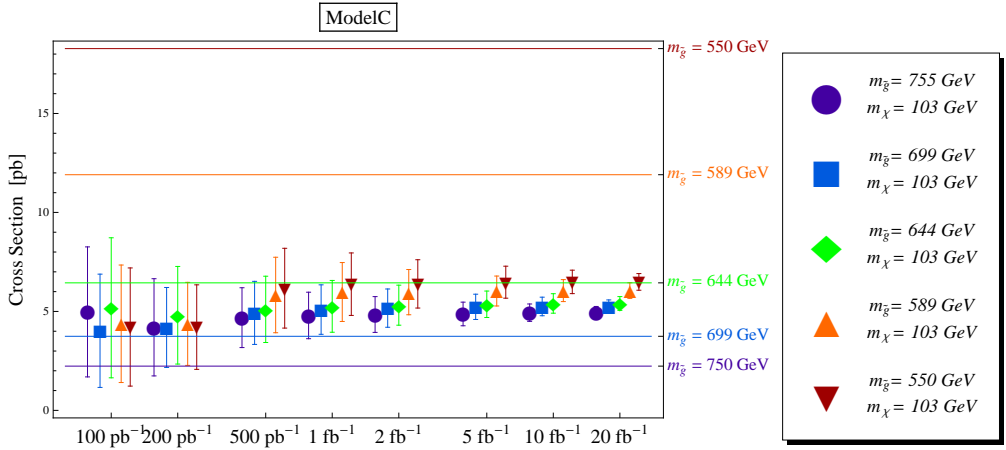}
\includegraphics[scale=0.35]{Legend.eps}
\caption{\label{crosssection}Gluino production cross section obtained by summing the
  individual branching ratios. We also show the theoretical cross
  sections at NNLO for gluino pair production for masses used in the
  templates. Notice that at high integrated luminosity we can begin to
  rule out heavier mass hypothesis}
\end{center}
\end{figure}

Notice that since our fit yield a set of values for
$\sqrt{\sigma_{\tilde{g}\tilde{g}}{\mathcal{L}}}\ Br_{{\rm
    channel}}$ with any given mass hypothesis. It can by itself
provide a consistency check on the hypothesis. In particular,  we can
get a lower limit on the gluino
pair production cross section, by summing and squaring the fit values
$\sqrt{\sigma_{\tilde{g}\tilde{g}}{\mathcal{L}}}\ Br_{{\rm
    channel}}$
\[\sigma_{\tilde{g}\tilde{g}}
\ge
\mathcal{L}^{-1}
\sum{(
\sqrt{\sigma_{\tilde{g}\tilde{g}}{\mathcal{L}}}Br_{tt}+
\sqrt{\sigma_{\tilde{g}\tilde{g}}{\mathcal{L}}}Br_{bb}+
\sqrt{\sigma_{\tilde{g}\tilde{g}}{\mathcal{L}}}Br_{tb})^2
}.\]
We can obtain cross sections for gluino
pair production in the case of
decoupled scalars, and rule out some incorrect gluino masses used in
the templates. For example, in Model C (see Figure
\ref{crosssection}), we can rule out the $m_{\tilde{g}}=755$ GeV mass
template at $200\mbox{ pb}^{-1}$ of data at $1\sigma$ certainty, and
begin to rule out the $m_{\tilde{g}}=700$ GeV mass template at
approximately $1\mbox{ fb}^{-1}$ of data at $1\sigma$ certainty. If the actual gluino is much lower
than 600 GeV then there must be a significant branching fraction of the gluino that is not
contributing the channels we used in our fits, such as gluino decays into first and second generation quarks.

\section{Conclusion}

We have studied the LHC signals of pair produced light gluinos which decay
dominantly through $\tilde{g} \rightarrow tt + \chi$, $\tilde{g}
\rightarrow bb + \chi$, and $\tilde{g} \rightarrow tb + \chi$. We
conclude that an early discovery of new physics in this scenario is
possible due to significant excesses expected in multi-lepton
multi-bottom channels. Measuring relative branching ratios of gluino
decay into $tt$, $bb$ and $tb$ channels is essential to extract
information about the underlying model. The crucial step in such a
measurement is identify top multiplicity in the signal events. We show
that direct reconstruction, while useful in gathering evidence for the
existence top quark in decay products, is not sufficient to measure
the number of top quarks in the event effectively. We proposed and studied
a method based on fitting a set of branching ratios to a collection of
experimental observables, most of them inclusive counts. Efficiencies
for identifying a particular final state resulting from certain
underlying decay topology are estimated by simulating corresponding
templates. We conclude that this method will allow us to learn about
gluino decay branching ratios with roughly 10 ${\rm fb}^{-1}$ of
integrated luminosity. We verified that, in combination with earlier
information on the mass spectrum of gluino and the LSP, we can obtain
a reliable measurement by choosing appropriate mass hypotheses. We
emphasize that the main advantage of our method is that it allows us
to use a large number of channels, many of them with multiple leptons
and bottoms, in which we expect to see excesses during the early stage of
LHC in models of this sort.

We showed that we can expect to gain enough information about the mass
spectrum, in particular the mass gap $M_{\tilde{g}}- M_{\rm LSP}$, from
simple observables like effective mass  and  demonstrated that, with
our fitting method, it is also possible to obtain an estimate of the
gluino production cross section which in turn gives us very valuable
information of the gluino mass, which also gives a consistency check
of the assumption we made in our measurement of the branching ratio.

In the paper, we have considered only benchmark models with only
a set of simple decay chains. More complicated models will certainly
contain channels which requires further study
\cite{Hisano:2002xq,Hisano:2003qu}.  For example, a
decay chain which contains $...\tilde{t}\rightarrow b\tilde{C}$ followed
by $\tilde{C}\rightarrow W\tilde{N}$ has the same set of final state
particles as top decay. On the other hand, we have also only used
counting signatures in our fit. Inclusion of  more kinematical
variables may improve our ability of discerning other decay
topologies. For example, in the decay of stop mentioned above, the
kinematics of $b$ and $W$ will be in general different from the case of
top decay.

One key factor entering our lepton efficiency is the isolation requirement.
This is particularly significant in our case since we expect to have
a lot of hadronic activity in these top rich events. We have
only implemented a simple and commonly used isolation criterion based
on hadronic activity within a narrow cone around the lepton. However,
one can in principle improve and optimize the isolation cuts by including
additional alternative isolation criteria, such as by requiring the
invariant mass of lepton and the hadronic activity $m_{\ell h}>m_{{\rm cut}}$.
This cut is effective since a lepton from heavy flavor decay is
typically soft and therefore gives a small $m_{\ell h}$, while it
is expected to be larger for accidental overlap between lepton and
jet.

\section{Acknowledgement}

We are grateful for interesting discussions with Aaron Pierce and David Morrissey. 
L.-T. W. is supported by the National
Science Foundation under grant PHY-0756966 and the Department of
Energy under grant DE-FG02-90ER40542. G.L.K., E.K. and P.G. are supported in part by 
the U.S. D.O.E. B.S.A and L.-T.W. thank the Michigan Center for
Theoretical Physics for hospitality during which this work was done.
G.L.K. thanks IAS for hospitality. E.K. thanks Princeton University for hospitality. 
P.G. thanks the Abdus Salam International Center for Theoretical Physics for hospitality.


\begin{thebibliography}{10}



\bibitem{top-composite}
 T.~Gherghetta and A.~Pomarol,
 Nucl.\ Phys.\  B {\bf 586}, 141 (2000)
 [arXiv:hep-ph/0003129].
 K.~Agashe, A.~Delgado, M.~J.~May and R.~Sundrum,
 JHEP {\bf 0308}, 050 (2003)
 [arXiv:hep-ph/0308036].


\bibitem{susy}
 S.~Dimopoulos and H.~Georgi,
 Nucl.\ Phys.\ B {\bf 193}, 150 (1981).
\bibitem{ued}
 T.~Appelquist, H.~C.~Cheng and B.~A.~Dobrescu,
 Phys.\ Rev.\  D {\bf 64}, 035002 (2001)
 [arXiv:hep-ph/0012100].
\bibitem{lh}
 N.~Arkani-Hamed, A.~G.~Cohen and H.~Georgi,
 Phys.\ Lett.\  B {\bf 513}, 232 (2001)
 [arXiv:hep-ph/0105239].
\bibitem{twin}
 Z.~Chacko, H.~S.~Goh and R.~Harnik,
 Phys.\ Rev.\ Lett.\  {\bf 96}, 231802 (2006)
 [arXiv:hep-ph/0506256].


\bibitem{resonance-ttbar}
K.~Agashe, A.~Belyaev, T.~Krupovnickas, G.~Perez and J.~Virzi,
 Phys.\ Rev.\  D {\bf 77}, 015003 (2008)
 [arXiv:hep-ph/0612015].
V.~Barger, T.~Han and D.~G.~E.~Walker,
 Phys.\ Rev.\ Lett.\  {\bf 100}, 031801 (2008)
 [arXiv:hep-ph/0612016].
 B.~Lillie, L.~Randall and L.~T.~Wang,
 JHEP {\bf 0709}, 074 (2007)
 [arXiv:hep-ph/0701166].
 A.~L.~Fitzpatrick, J.~Kaplan, L.~Randall and L.~T.~Wang,
 JHEP {\bf 0709}, 013 (2007)
 [arXiv:hep-ph/0701150].
 K.~Agashe, H.~Davoudiasl, G.~Perez and A.~Soni,
 Phys.\ Rev.\  D {\bf 76}, 036006 (2007)
 [arXiv:hep-ph/0701186].
 U.~Baur and L.~H.~Orr,
 Phys.\ Rev.\  D {\bf 76}, 094012 (2007)
 [arXiv:0707.2066 [hep-ph]].
 R.~Frederix and F.~Maltoni,
 arXiv:0712.2355 [hep-ph].
 U.~Baur and L.~H.~Orr,
 arXiv:0803.1160 [hep-ph].
 J.~Thaler and L.~T.~Wang,
  JHEP {\bf 0807}, 092 (2008)
  [arXiv:0806.0023 [hep-ph]].
D.~E.~Kaplan, K.~Rehermann, M.~D.~Schwartz and B.~Tweedie,
  Phys.\ Rev.\ Lett.\  {\bf 101}, 142001 (2008)
  [arXiv:0806.0848 [hep-ph]].
L.~G.~Almeida, S.~J.~Lee, G.~Perez, G.~Sterman, I.~Sung and J.~Virzi,
  arXiv:0807.0234 [hep-ph].
L.~G.~Almeida, S.~J.~Lee, G.~Perez, I.~Sung and J.~Virzi,
  arXiv:0810.0934 [hep-ph].
\bibitem{Brooijmans}
G. Brooijmans, ATLAS note, ATL-PHYS-CONF-2008-008.

\bibitem{Bai:2008sk}
  Y.~Bai and Z.~Han,
  arXiv:0809.4487 [hep-ph].

\bibitem{tprime-ttet}
 H.~C.~Cheng, I.~Low and L.~T.~Wang,
 Phys.\ Rev.\ D {\bf 74}, 055001 (2006)
 [arXiv:hep-ph/0510225].
 P.~Meade and M.~Reece,
 Phys.\ Rev.\ D {\bf 74}, 015010 (2006)
 [arXiv:hep-ph/0601124].
A.~Freitas and D.~Wyler,
 JHEP {\bf 0611}, 061 (2006)
 [arXiv:hep-ph/0609103].
   A.~Belyaev, C.~R.~Chen, K.~Tobe and C.~P.~Yuan,
 arXiv:hep-ph/0609179.
S.~Matsumoto, M.~M.~Nojiri and D.~Nomura,
 Phys.\ Rev.\  D {\bf 75}, 055006 (2007)
 [arXiv:hep-ph/0612249];
 M.~M.~Nojiri and M.~Takeuchi,
 arXiv:0802.4142 [hep-ph].
T.~Han, R.~Mahbubani, D.~G.~E.~Walker and L.~T.~Wang,
  arXiv:0803.3820 [hep-ph].

\bibitem{Contino:2008hi}
  R.~Contino and G.~Servant,
  JHEP {\bf 0806}, 026 (2008)
  [arXiv:0801.1679 [hep-ph]].



\bibitem{Kraml:2005kb} S.~Kraml and A.~R.~Raklev, 

\bibitem{Lillie:2007hd}
  B.~Lillie, J.~Shu and T.~M.~P.~Tait,
  JHEP {\bf 0804}, 087 (2008)
  [arXiv:0712.3057 [hep-ph]].

\bibitem{Han:2008xb}
  For a recent review on top quark related new physics signatures, see
  T.~Han,
  arXiv:0804.3178 [hep-ph].



\bibitem{Chung:2003fi}
For a recent review, see
  D.~J.~H.~Chung, L.~L.~Everett, G.~L.~Kane, S.~F.~King, J.~D.~Lykken
  and L.~T.~Wang,
  Phys.\ Rept.\  {\bf 407}, 1 (2005)
  [arXiv:hep-ph/0312378].

\bibitem{Gerbush:2007fe}
  M.~Gerbush, T.~J.~Khoo, D.~J.~Phalen, A.~Pierce and D.~Tucker-Smith,
  Phys.\ Rev.\  D {\bf 77}, 095003 (2008)
  [arXiv:0710.3133 [hep-ph]].
\bibitem{scalar-octet}
  A.~R.~Zerwekh, C.~O.~Dib and R.~Rosenfeld,
  Phys.\ Rev.\  D {\bf 77}, 097703 (2008)
  [arXiv:0802.4303 [hep-ph]].
  P.~Fileviez Perez, R.~Gavin, T.~McElmurry and F.~Petriello,
  arXiv:0809.2106 [hep-ph].
  T.~Plehn and T.~M.~P.~Tait,
  arXiv:0810.3919 [hep-ph].
\bibitem{Dobrescu:2007yp}
  B.~A.~Dobrescu, K.~Kong and R.~Mahbubani,
  arXiv:0709.2378 [hep-ph].


\bibitem{zprime-med}
 P.~Langacker, G.~Paz, L.~T.~Wang and I.~Yavin,
  Phys.\ Rev.\ Lett.\  {\bf 100}, 041802 (2008)
  [arXiv:0710.1632 [hep-ph]].
H.~Verlinde, L.~T.~Wang, M.~Wijnholt and I.~Yavin,
  JHEP {\bf 0802}, 082 (2008)
  [arXiv:0711.3214 [hep-th]].
  P.~Langacker, G.~Paz, L.~T.~Wang and I.~Yavin,
  Phys.\ Rev.\  D {\bf 77}, 085033 (2008)
  [arXiv:0801.3693 [hep-ph]].
\bibitem{mirage}
  L.~L.~Everett, I.~W.~Kim, P.~Ouyang and K.~M.~Zurek,
  JHEP {\bf 0808}, 102 (2008)
  [arXiv:0806.2330 [hep-ph]].
  L.~L.~Everett, I.~W.~Kim, P.~Ouyang and K.~M.~Zurek,
  Phys.\ Rev.\ Lett.\  {\bf 101}, 101803 (2008)
  [arXiv:0804.0592 [hep-ph]].
  
  S.~Nakamura, K.~i.~Okumura and M.~Yamaguchi,
  Phys.\ Rev.\  D {\bf 77}, 115027 (2008)
  [arXiv:0803.3725 [hep-ph]].

  K.~Choi, K.~S.~Jeong, S.~Nakamura, K.~I.~Okumura and M.~Yamaguchi,
  arXiv:0901.0052 [hep-ph].


\bibitem{Acharya:2008zi}
  B.~S.~Acharya, K.~Bobkov, G.~L.~Kane, J.~Shao and P.~Kumar,
  arXiv:0801.0478 [hep-ph].

\bibitem{Heckman:2008rb}
  J.~J.~Heckman and C.~Vafa,
  arXiv:0809.3452 [hep-ph].
 

\bibitem{eff_susy}
 A.~G.~Cohen, D.~B.~Kaplan, F.~Lepeintre and A.~E.~Nelson,
 Phys.\ Rev.\ Lett.\  {\bf 78} (1997) 2300
 [arXiv:hep-ph/9610252].

\bibitem{focus_point}
 J.~L.~Feng, K.~T.~Matchev and T.~Moroi,
 Phys.\ Rev.\  D {\bf 61} (2000) 075005
 [arXiv:hep-ph/9909334].

\bibitem{split}
N.~Arkani-Hamed and S.~Dimopoulos,
 JHEP {\bf 0506} (2005) 073
 [arXiv:hep-th/0405159].
 G.~F.~Giudice and A.~Romanino,
 Nucl.\ Phys.\  B {\bf 699} (2004) 65
 [Erratum-ibid.\  B {\bf 706} (2005) 65]
 [arXiv:hep-ph/0406088].
 N.~Arkani-Hamed, S.~Dimopoulos, G.~F.~Giudice and A.~Romanino,
 Nucl.\ Phys.\  B {\bf 709} (2005) 3
 [arXiv:hep-ph/0409232].



\bibitem{Baer:1990sc}
H.~Baer, X.~Tata and J.~Woodside,
  ``PHENOMENOLOGY OF GLUINO DECAYS VIA LOOPS AND TOP QUARK YUKAWA
  COUPLING,''   Phys.\ Rev.\  D {\bf 42}, 1568 (1990).   



\bibitem{Hisano:2002xq} J.~Hisano, K.~Kawagoe, R.~Kitano and M.~M.~Nojiri,
``Scenery from the top: Study of the third generation squarks at CERN
  LHC,''   Phys.\ Rev.\  D {\bf 66}, 115004 (2002)
  [arXiv:hep-ph/0204078].   





\bibitem{Hisano:2003qu} J.~Hisano, K.~Kawagoe and M.~M.~Nojiri,
``A detailed study of the gluino decay into the third generation
  squarks  at   


\bibitem{Mercadante:2007zz}
  P.~G.~Mercadante, J.~K.~Mizukoshi and X.~Tata,
  Braz.\ J.\ Phys.\  {\bf 37}, 549 (2007).



\bibitem{Baer:2007ya}
  H.~Baer, V.~Barger, G.~Shaughnessy, H.~Summy and L.~t.~Wang,
  Phys.\ Rev.\  D {\bf 75}, 095010 (2007)
  [arXiv:hep-ph/0703289].



\bibitem{Gambino:2005eh} P.~Gambino, G.~F.~Giudice and P.~Slavich,
``Gluino decays in split supersymmetry,''   Nucl.\ Phys.\  B {\bf
  726}, 35 (2005)   [arXiv:hep-ph/0506214].   



\bibitem{Toharia:2005gm} M.~Toharia and J.~D.~Wells, ``Gluino decays
  with heavier scalar superpartners,''   JHEP {\bf 0602}, 015 (2006)
  [arXiv:hep-ph/0503175].   






\bibitem{Bonciani:1998vc}  R.~Bonciani, S.~Catani, M.~L.~Mangano and P.~Nason,
  ``NLL resummation of the heavy-quark hadroproduction cross-section,''
  Nucl.\ Phys.\  B {\bf 529}, 424 (1998)
  [arXiv:hep-ph/9801375].

\bibitem{Sjostrand:2006za} T.~Sjostrand, S.~Mrenna and P.~Skands,
``PYTHIA 6.4 physics and manual,''   JHEP {\bf 0605}, 026 (2006)
  [arXiv:hep-ph/0603175].   

\bibitem{Alwall:2007st} J.~Alwall \textit{et al.}, ``MadGraph/MadEvent
  v4: The New Web Generation,''   JHEP {\bf 0709}, 028 (2007)
  [arXiv:0706.2334 [hep-ph]].   

\bibitem{Catani:2001cc}
  S.~Catani, F.~Krauss, R.~Kuhn and B.~R.~Webber,
  JHEP {\bf 0111}, 063 (2001)
  [arXiv:hep-ph/0109231].

\bibitem{PGS4}John Conway,
  ``http://www.physics.ucdavis.edu/~conway/research/software/pgs/pgs4-general.htm'



\bibitem{Beenakker:1996ch}
  W.~Beenakker, R.~Hopker, M.~Spira and P.~M.~Zerwas,
  Nucl.\ Phys.\  B {\bf 492}, 51 (1997)
  [arXiv:hep-ph/9610490].

\bibitem{samesign_dilepton}
See, for example,   H.~Baer, C.~h.~Chen, F.~Paige and X.~Tata,
 Phys.\ Rev.\  D {\bf 53} (1996) 6241
 [arXiv:hep-ph/9512383].


\bibitem{Baer:2008kc}
  H.~Baer, H.~Prosper and H.~Summy,
  Phys.\ Rev.\  D {\bf 77}, 055017 (2008)
  [arXiv:0801.3799 [hep-ph]].
  
\bibitem{Sullivan:2008ki}
  Z.~Sullivan and E.~L.~Berger,
  Phys.\ Rev.\  D {\bf 78}, 034030 (2008)
  [arXiv:0805.3720 [hep-ph]].




\bibitem{Acharya:2008tz} B.~S.~Acharya, F.~Cavallari, G.~Corcella,
R.~Di Sipio and G.~Petrucciani, ``Re-discovery of the top quark at the
LHC and first measurements,''   arXiv:0806.0484 [hep-ex].   



\bibitem{Nojiri:2003tv}
  M.~M.~Nojiri, G.~Polesello and D.~R.~Tovey,
  arXiv:hep-ph/0312318.


\bibitem{ArkaniHamed:2007fw}
  N.~Arkani-Hamed, P.~Schuster, N.~Toro, J.~Thaler, L.~T.~Wang,
  B.~Knuteson and S.~Mrenna,
  arXiv:hep-ph/0703088.

\end{thebibliography}
\end{document}